\documentclass[twocolumn]{aastex631}

\usepackage{array,multirow}

\shorttitle{Kinematics of molecular clouds}
\shortauthors{Yuan et al.}
\graphicspath{{./}{figures/}}

\begin{document}

\title{Cloud-to-cloud velocity dispersions across a Local arm segment}

\correspondingauthor{Ji Yang}
\email{jiyang@pmo.ac.cn}

\author[0000-0003-0804-9055]{Lixia Yuan}
\affiliation{Purple Mountain Observatory and Key Laboratory of Radio Astronomy, Chinese Academy of Sciences, \\
10 Yuanhua Road, Qixia District, Nanjing 210033, PR China}

\author[0000-0001-7768-7320]{Ji Yang}
\affiliation{Purple Mountain Observatory and Key Laboratory of Radio Astronomy, Chinese Academy of Sciences, \\
10 Yuanhua Road, Qixia District, Nanjing 210033, PR China}

\begin{abstract}

Using a large sample of 9617 molecular clouds (MCs) from the Milky Way Imaging Scroll Painting survey, 
we mainly measure one-dimensional cloud-to-cloud velocity dispersions across a 450 deg$^{2}$ segment of the Local arm in the Galactic second quadrant. 
We define the cloud-to-cloud velocity dispersion using two metrics: the standard deviation ($\sigma_{\rm bin}$) 
and flux-weighted root-mean-squared value ($\sigma_{\rm bin, w}$) of the centroid velocities of $^{12}$CO-detected MCs 
within spatial bins. 
The typical values of $\sigma_{\rm bin}$ and $\sigma_{\rm bin, w}$ are 7.5$\pm$0.5 km s$^{-1}$ and 6.2$\pm$0.5 km s$^{-1}$, respectively. 
After categorizing clouds by sizes into three types: 
Type $S$ (0.15 -- 1.2 pc), Type $M$ (1.2 -- 4.8 pc), and Type L ($\gtrsim$ 4.8 pc), 
we find that the spatial distribution of Type $S$ and $M$ MCs 
projected onto the Galactic longitude-latitude (l-b) plane is generally uniform.
Additionally, the cloud-to-cloud velocity dispersion among Type $S$ clouds 
($\sim$ 7.6 and 7.4 km s$^{-1}$ for $\sigma_{\rm bin}$ and $\sigma_{\rm bin, w}$, respectively) 
is systematically greater than that among Type $M$ clouds 
($\sim$ 6.7 and 6.0 km s$^{-1}$ for $\sigma_{\rm bin}$ and $\sigma_{\rm bin, w}$, respectively), 
with differences of 0.9 -- 1.4 km s$^{-1}$. 
From these measurements, we estimate merger timescales between MCs to be approximately 0.3 to 0.9 Myr, 
which is shorter than their internal crossing timescales ($\sim$ 1 Myr for Type $S$, $\sim$ 2 Myr for Type $M$, 
and $\gtrsim$ 5 Myr for Type $L$). 
This disparity, particularly pronounced for larger Type $L$ clouds, 
suggests that MCs are dynamically transient structures, 
with their gas content changing due to frequent interactions with neighboring clouds.

\end{abstract}


\section{Introduction}\label{sec:intro}

Molecular clouds (MCs) exist in a dynamic interstellar environment where they are not isolated systems. 
They could undergo gas flows in and out of the atomic phase, as well as mergers and splits with other MCs. 
These processes involve the exchange of mass, energy, and momentum.
\citep{Tasker2009, Dobbs2011, Dobbs2013, Dobbs2015, Fukui2016, Gong2017, Fukui2018, Jeffreson2021, Skarbinski2023, Jeffreson2024}. 
Inside MCs, smaller and denser substructures exist at multiple hierarchical levels within MCs. 
\citep{Larson1981, Myers1983, McKee1989, Ballesteros1999, MacLow2004, Molinari2010, Andre2010, Andre2014, Yuan2020}. 
However, external interactions among MCs most likely leave imprints on these inner substructures. 
Therefore, studying the distribution and movement of these substructures can provide significant insights into 
the assembly history of MCs \citep{Takahira2018, Fukui2021}. 

In a series of studies using a large sample of MCs from the Milky Way Imaging Scroll Painting (MWISP) CO survey, 
we conducted a systematic analysis of $^{13}$CO structures within MCs, 
focusing on their content, spatial distribution, and relative motion. 
Our findings indicate that, as MCs increase in scale, 
they contain a greater number of internal $^{13}$CO structures \citep[hereafter Paper II]{Yuan2022}. 
These structures tend to be regularly spaced with a preferred separation \citep[hereafter Paper III]{Yuan2023}. 
Additionally, these structures exhibit random relative motions \citep[hereafter Paper V]{Yuan2024}, 
which dominate the internal velocity dispersion of MCs \citep[hereafter Paper IV]{Yuan2023b}. 
Based on these observed results, we proposed an alternative picture for the assembly and destruction of MCs: 
regularly spaced $^{13}$CO structures serve as the building blocks of MCs, while transient processes of 
MCs occur through slow mergers or splits among these fundamental blocks. 
Notably, these processes may preserve the density structure of MCs while influencing their overall velocity.  
Furthermore, approximately 40$\%$ of MCs containing double or triple $^{13}$CO structures 
are observed to exhibit distinct velocity discontinuities between $^{13}$CO structures \citep{Yuan2024}. 
This leads us to question whether the velocity fields within MCs inherit the velocity fields between clouds, 
indicating mergers or splits among the `fundamental blocks' of MCs. 
Understanding cloud-to-cloud motion is crucial for uncovering the connections between 
inter-cloud and intra-clouds velocity fields. 
Additionally, cloud-to-cloud motion is directly related to the merger rates between MCs, 
which determines the frequency of interactions among these clouds.  

In observation, cloud-to-cloud velocity dispersion ($\sigma_{\rm cl}$) is a fundamental metric for measuring the cloud dynamics. 
Previous studies have measured this dispersion across various regions of the Galaxy using different CO surveys. 
In the solar neighborhood ($\lesssim$ 3 kpc), the one-dimensional velocity dispersion for MCs 
with masses ranging from 10$^{2}$ -- 10$^{4}$ M$_{\odot}$ is 9.0$^{+1.0}_{-1.1}$ km s$^{-1}$.
For more massive clouds (10$^{4}$ -- 10$^{5.5}$ M$_{\odot}$), 
the velocity dispersion is about 6.6$^{+0.9}_{-0.6}$ km s$^{-1}$. 
Using a larger database, 
this value has been recalculated as 7.8$^{+0.6}_{-0.5}$ km s$^{-1}$ for masses 
typically 0.5 to 5$\times$10$^{5}$ M$_{\odot}$ \citep{Stark1989}. 
\cite{Blitz1984} reported a $\sigma_{\rm cl}$ value of 5.7$\pm$1.2 km s$^{-1}$ for high-latitude local MCs with masses roughly 10$^{2}$ M$_{\odot}$. 
Across the Galaxy, \cite{Clemens1985} estimated the $\sigma_{\rm cl}$ values to 
range from 2 km s$^{-1}$ and 8 km s$^{-1}$ for distances between 2 kpc and 10 kpc. 
These estimates were derived by fitting Gaussian functions to CO cloud velocities at tangent points 
using data from Massachusetts-Stony Brook Galactic plane CO survey. 
More recently, \cite{Su2021} analyzed MCs at the tangent points using data from the MWISP CO survey, 
which offers higher sensitivity and resolution. 
They found the $\sigma_{\rm cl}$ value is 4.9$\pm$1.3 km s$^{-1}$, 
observing a linear variation with a slope of $\sim$ -0.4 km s$^{-1}$ kpc$^{-1}$ along the Galactic distance from 2.2 to 6.4 kpc. 
In nearby galaxies, \cite{wilson2011} employed CO(3-2) spectral lines observed by James Clerk Maxwell Telescope (JCMT) 
for a sample of twelve nearby spiral galaxies. They found the average $\sigma_{\rm cl}$ values across the galactic discs is 
6.1$\pm$1.0 km s$^{-1}$. However, it is important to note that measurements of cloud-to-cloud velocity dispersions can be influenced by the resolution and completeness of 
observed molecular gas, which are often limited by the resolution and sensitivity of CO spectral line observations. 
A more extensive and detailed dataset could provide new insights into the formation and dynamics of clouds, 
particularly for smaller-scale and faint clouds that can easily be diluted in previous observations.

In this paper, we present a systematic analysis of cloud-to-cloud velocity dispersions and cloud number distributions, 
using a larger sample of MCs identified in the MWISP CO survey, which has higher sensitivity 
and resolution (spectral and spatial). 
The structure of this paper is organized as follows: 
Section 2 describes the $^{12}$CO line data from the MWISP CO survey and the extracted $^{12}$CO molecular clouds. 
In Section 3, we present the results of measured cloud-to-cloud velocity dispersions and cloud number distributions.
Section 4 discusses cloud-to-cloud velocity dispersions across Galactic segments with different distances, 
estimates merger rates among MCs, and examines their implications for the dynamical evolution of MCs. 
Finally, Section 5 summarizes our findings.

\section{Data}

\subsection{$^{12}$CO (J=1-0) spectral-line data from the Milky Way Imaging Scroll Painting Survey} 
The MWISP survey is an ongoing survey of carbon monoxide (CO) emissions in the northern Galactic plane, 
conducted using the 13.7m telescope located in Delingha, China. 
This survey simultaneously observes the $^{12}$CO, $^{13}$CO, and C$^{18}$O at $J$ = 1-0 transition lines 
utilizing the full-sampling On-The-Fly mapping techniques \citep{sun2018}. 
A detailed description of the telescope's performance and its multibeam receiver system can be found in \cite{Su2019, Shan2012}. 
Additionally, the observational strategy and raw data reduction processes used in the MWISP survey are outlined in \cite{Su2019}. 
The half-power beamwidth (HPBW) of the antenna at a frequency of 115 GHz is $\sim$ 50$^{\prime \prime}$. 
The typical system temperature is $\sim$ 250 K for $^{12}$CO lines (115.271 GHz) in the upper sideband and 
$\sim$ 140 K for the $^{13}$CO (110.201 GHz) and C$^{18}$O (109.782 GHz) lines in the lower sideband, respectively. 
This survey offers a total bandwidth of 1 GHz with 16,384 channels, providing a spectral resolution of 61 kHz per channel. 
This results in a velocity resolution of approximately 0.16 km s$^{-1}$ for $^{12}$CO lines and 0.17 km s$^{-1}$ for $^{13}$CO and C$^{18}$O lines.  
The typical root mean square (RMS) noise levels achieved for the $^{12}$CO and $^{13}$CO lines are $\sim$ 0.5 K and $\sim$ 0.3 K, respectively. 
Additionally, \cite{Jiang2023} present the distribution of main-beam temperatures (T$_{\rm MB}$) and radial velocities 
concerning the local standard of rest (V$_{\rm LSR}$) of a `standard source' L134 over a seven-year period 
from 2011 to 2017. During this period, the 1$\sigma$ variation of T$_{\rm MB}$ of the $^{12}$CO lines was 0.66 K, 
which corresponds to a relative change of 7.4$\%$. 
The velocity shift between $^{12}$CO and $^{13}$CO lines is approximately $\sim$ 0.15 km s$^{-1}$. 
These results suggest the observation system remained reasonably stable throughout this period.

In this study, we analyze $^{12}$CO lines data from the MWISP survey, covering a region of 450 deg$^{2}$ 
with Galactical longitude $l$ ranging from 104$^{\circ}$.75 to 150$^{\circ}$.25, 
Galactical latitude $|b| < 5^{\circ}.25$, and the line-of-sight velocity of $-$95 km s$^{-1}$  $<$ V$_{\rm LSR}$ $<$ 25 km s$^{-1}$.  
These $^{12}$CO lines emission data have also been analyzed in our previous series of works in \cite{Yuan2021, Yuan2022, Yuan2023, Yuan2023b, Yuan2024}. 

\subsection{The extracted $^{12}$CO molecular cloud samples \label{sec:MC_samples}}
In our analysis, the $^{12}$CO molecular cloud is defined as a collection of adjacent voxels within 
the position-position-velocity (PPV) space where observed $^{12}$CO(1-0) line intensities above a certain threshold. 
We employed the Density-based Spatial Clustering of Applications with Noise (DBSCAN) algorithm, 
which is designed to identify clusters with arbitrary shapes in large spatial databases \citep{ester1996}, 
to extract MCs in the $^{12}$CO data cube, as described by \cite{Yan2020}.
This algorithm takes into account both intensity levels and signal continuity, 
making it suitable for the extended and irregular shapes of MCs. 
Three parameters are utilized in the DBSCAN algorithm to extract the $^{12}$CO MCs. 
The first parameter, \textit{cutoff}, determines the line intensity threshold. 
The other two parameters, $\epsilon$ and \textit{MinPts}, define the connectivity of the identified structures. 
A core point within extracted structures is characterized by having its adjacent points within a specified radius ($\epsilon$) 
exceeding a threshold number (\textit{MinPts}). 
A border point, on the other hand, is located inside the $\epsilon$-radius of a core point, 
but its adjacent points within the radius of $\epsilon$ fails to meet the \textit{MinPts} requirement \citep{Yan2020}.  

For the identification of $^{12}$CO clouds, we used the following parameters: 
\textit{cutoff}=2$\sigma$, where $\sigma$ is the rms noise,  approximately 0.5 K for the $^{12}$CO line emission, 
\textit{MinPts}=4, and $\epsilon$=1. These values were suggested by \citep{Yan2020}.  
Additionally, we applied post-selection criteria to minimize noise contamination, 
which includes: (1) the total number of voxels in each extracted structure must be greater than 16; 
(2) the peak intensity of extracted voxels must exceed the \textit{cutoff} value plus 3$\sigma$;
(3) the angular area of the extracted structure must be larger than one beam size (2$\times$2 pixels, approximately 1$^{\prime}$);
and (4) the number of velocity channels must exceed 3. 
Using these parameters and criteria, a catalog of 18,190 $^{12}$CO molecular clouds was identified in the above region by the DBSCAN algorithm \citep{Yan2021}.  
We visually inspected and classified these 18,190 MCs into filaments and nonfilaments in \citep[hereafter Paper I]{Yuan2021}. 
Additionally, the dependence of extracted MC samples on finite angular resolution, 
spectral line sensitivity, and different algorithms has been systematically investigated in \cite{Yan2022}. 
The extracted $^{12}$CO line data for the 18,190 $^{12}$CO clouds 
have been publicly available at DOI:\href{https://doi.org/10.57760/sciencedb.j00001.00427}{10.57760/sciencedb.j00001.00427}.
It is important to note that the 18,190 MCs were reported in \cite{Yan2021, Yuan2021}. 
At that time, certain subregions (angular areas $<$ 5 deg$^{2}$), mainly located within l=(120$^{\circ}$, 130$^{\circ}$) $\&$ b=(-5$^{\circ}$.25, -3$^{\circ}$.75), 
l=(129$^{\circ}$.25, 129$^{\circ}$.75) $\&$ b=(4$^{\circ}$.75, 5$^{\circ}$.25), and l=(149$^{\circ}$.5, 150$^{\circ}$.25) $\&$ b=(-5$^{\circ}$.25, -4$^{\circ}$.25), 
were not fully surveyed, as indicated by the blank areas shown in Figure A2 of \cite{Yuan2021}. 
In this study, we apply the same DBSCAN parameters to extract MC samples from the complete 450 deg$^{2}$ region, 
and supplement MC samples extracted in the previously unobserved subregions. 
As a result, the updated MC catalog contains a total of 18301 MCs within the completed region defined by l=(105$^{\circ}$, 150$^{\circ}$) $\&$ b=(-5$^{\circ}$.25, 5$^{\circ}$.25).

\section{Result} 

\subsection{Cloud-to-cloud velocity dispersions defined in the specific bins \label{sec:MC_whole}}

The cloud-to-cloud velocity dispersion is related to the positions, distances, and velocities of MCs. 
Our observed MCs are identified as contiguous structures in position-position-velocity (PPV) space, 
with their positions projected onto the Galactic l-b plane and their velocities along the line of sight. 
According to the spiral structure model of the Milky Way \citep{Reid2016, Reid2019, Xu2023}, 
we define a total of 9617 MC samples with centroid velocities ranging from -30 to 25 km s$^{-1}$ 
as the `Near' region. These MCs are primarily located in the Local arm 
and have a typical kinematical distance of $\sim$ 0.5 kpc. 
We focus on MC samples in the `Near' region, which helps constrain their distances.  
The entire area of a 450 deg$^{2}$ region is divided into uniform bins. 
Within each bin, the extracted $^{12}$CO MCs are most likely to be close to each other, 
which further constrains their positions. 

We employ two methods to calculate the one-dimensional cloud-to-cloud velocity dispersion in each bin. 
One method utilizes the integrated fluxes of MCs as weights, 
calculating the root-mean-squared (rms) centroid velocities of $^{12}$CO clouds 
weighted by their integrated fluxes, denoted as $\sigma_{\rm bin, w}$. 
The other method assigns equal weight to the whole clouds, 
calculating the standard deviation of the centroid velocities of $^{12}$CO clouds.    
In a given bin, if the number of $^{12}$CO clouds is $N$, 
we calculate the cloud-to-cloud velocity dispersion as follows:

\begin{eqnarray} \label{e:sigma_lsr}
    \sigma_{\rm bin, w}^{2} = \Sigma^{\rm bin}_{j} F_{\rm j} (V_{\rm cen, j}-\overline{V}_{\rm cen, w})^{2}/\Sigma^{\rm bin}_{j} F_{\rm j}, \nonumber \\
    \overline{V}_{\rm cen, w}= \Sigma^{\rm bin}_{j} F_{\rm j} V_{\rm cen, j}/\Sigma^{\rm bin}_{j} F_{\rm j}, \nonumber \\
    \sigma_{\rm bin}^{2} = \Sigma^{\rm bin}_{j} (V_{\rm cen, j}-\overline{V}_{\rm cen})^{2}/(N-1), \nonumber \\
    \overline{V}_{\rm cen}= \Sigma^{\rm bin}_{j} V_{\rm cen, j}/N, \nonumber \\
\end{eqnarray}

where the sum $\Sigma^{\rm bin}_{i}$ runs over the whole $^{12}$CO clouds within a given bin, 
$V_{\rm cen, j}$ = $\Sigma^{\rm cloud}_{j}T_{ij}V_{ij}/\Sigma^{\rm cloud}_{j}T_{ij}$ 
is defined as the centroid velocity of the $j$th $^{12}$CO cloud, 
the integrated flux of the $j$th $^{12}$CO MC is given by 
$F_{j}=\int T_{mb}(l,b,v)dldbdv=0.167 \times 0.25 \Sigma_{i}^{jth} T_{\rm ji}$ (K km s$^{-1}$ arcmin$^{2}$).   
The $T_{ji}$ and $V_{ji}$ are the brightness temperature and the line-of-sight velocity of $i$th voxel in the $j$th $^{12}$CO cloud.  
The flux-weighted centroid velocity across clouds in the bin is denoted as $\overline{V}_{\rm cen, w}$, 
and the flux-weighted cloud-to-cloud velocity dispersion is represented by $\sigma_{\rm bin, w}$. 
The $\overline{V}_{\rm cen}$ indicates the mean of the centroid velocities of $^{12}$CO clouds,  
while $\sigma_{\rm bin}$ stands for the standard deviation of these centroid velocities of $^{12}$CO clouds within the bin.   

\begin{figure*}[ht]
    \centering
    \includegraphics[width=1.0\textwidth, trim=70 70 200 70, clip]{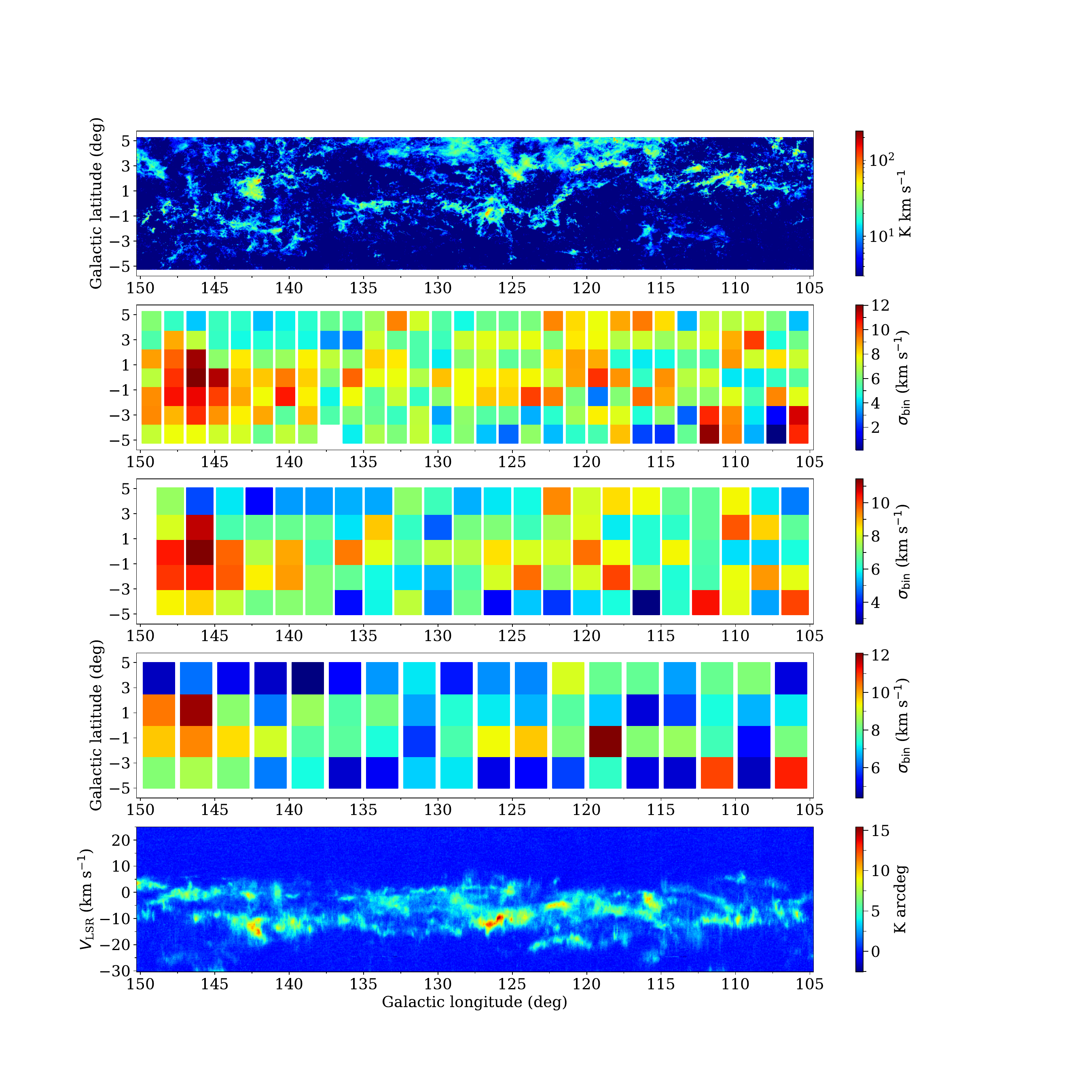} 
    \caption{\textbf{Upper panel}: the velocity-integrated map of $^{12}$CO(1-0) line emission in the velocity range of 
    (-30 25) km s$^{-1}$ with the threshold of 3 K km s$^{-1}$. \textbf{Middle three panels}: the distribution of 
    cloud-to-cloud velocity dispersion ($\sigma_{\rm bin}$), calculated at the bin sizes of 1.5$^{\circ}$, 2$^{\circ}$ and 2.5$^{\circ}$, respectively. 
    \textbf{Lower panel}: the longtitude-velocity map of $^{12}$CO(1-0) emission, 
    which is integrated along the Galactic latitude of (-5.25 5.25) deg. \label{fig:cl_map_near}}
\end{figure*}

Figure \ref{fig:cl_map_near} presents the distributions of the cloud-to-cloud velocity dispersion ($\sigma_{\rm bin}$) in each bin. 
We show the values calculated for bin areas of $1.5^{\circ} \times 1.5^{\circ}$, $2^{\circ} \times 2^{\circ}$, 
and $2.5^{\circ} \times 2.5^{\circ}$, respectively.
For comparison, we also display the velocity-integrated intensity and latitude-integrated intensity maps of the $^{12}$CO(1-0) emission. 
We find that the $\sigma_{\rm bin, w}$ values of bins predominantly range from $\sim$ 5 - 8 km s$^{-1}$. 
However, several bins, such as those in the ranges $l$= 111$^{\circ}$ - 113$^{\circ}$ and 117$^{\circ}$ - 120$^{\circ}$, 
exhibit $\sigma_{\rm bin}$ values close to $\sim$ 10 km s$^{-1}$. 
From the bottom l-v map, we observe gas streams with diverging velocities of approximately 10 km s$^{-1}$ in these $l$ ranges, 
which may contribute to the higher $\sigma_{\rm bin}$ observed. 

To quantify the effects of bin sizes and statistic errors on the calculated cloud-to-cloud velocity dispersion ($\sigma_{\rm bin}$ and $\sigma_{\rm bin, w}$), 
we systematically calculate the $\sigma_{\rm bin}$ and $\sigma_{\rm bin, w}$ in the uniform bins with varying sizes. 
To ensure the statistical significance, we aim for a minimum of $\sim$ 10 clouds in each bin, 
thus the smallest bin size in the near region is set at 2$^{\circ}$. 
Additionally, we gradually increase the bin sizes from 2$^{\circ}$ to 5$^{\circ}$ in increments of 0.5$^{\circ}$, 
to further assess the effects of different bin sizes on the $\sigma_{\rm bin}$ and $\sigma_{\rm bin, w}$ values.
For a typical distance of 0.5 kpc in the `Near' region, an angular size of 1$^{\circ}$ corresponds to 0.9 pc, 
indicating that the detection threshold for cloud density is about 3 pc$^{-2}$. 

\begin{figure*}[ht]
    \centering
    \includegraphics[width=1.0\textwidth, trim=120 30 120 30, clip]{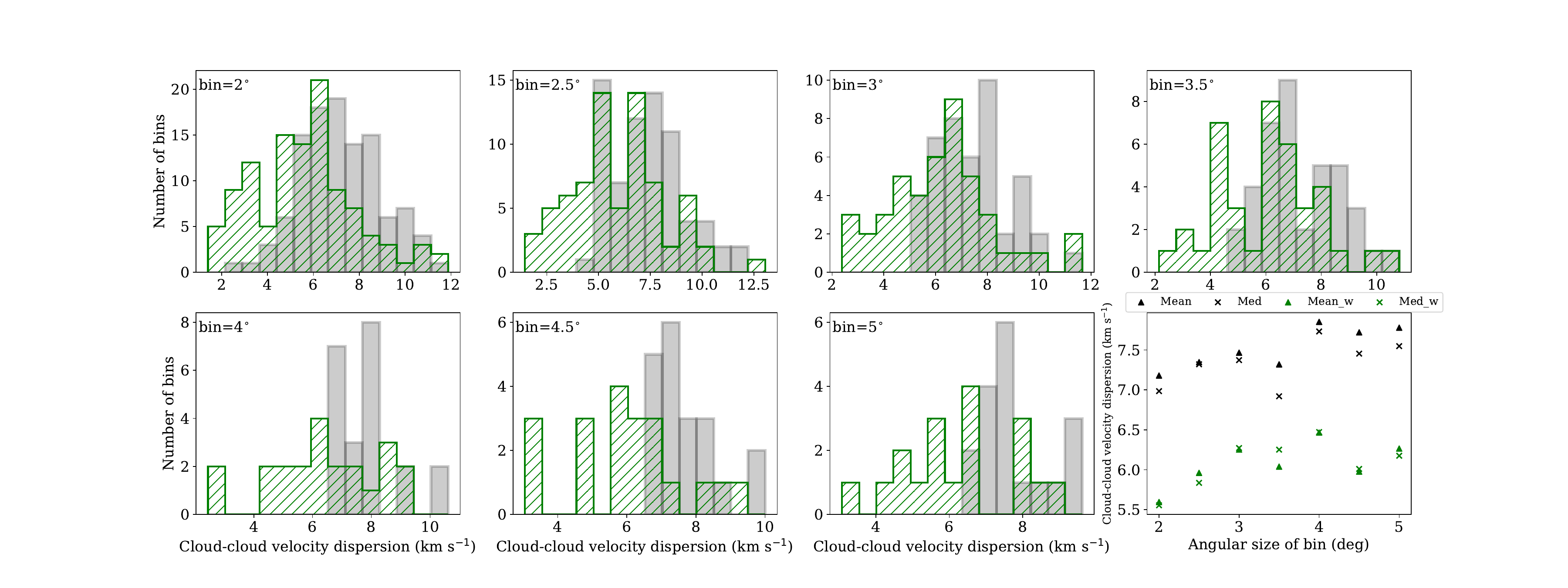}
    \includegraphics[width=1.0\textwidth, trim=120 30 120 30, clip]{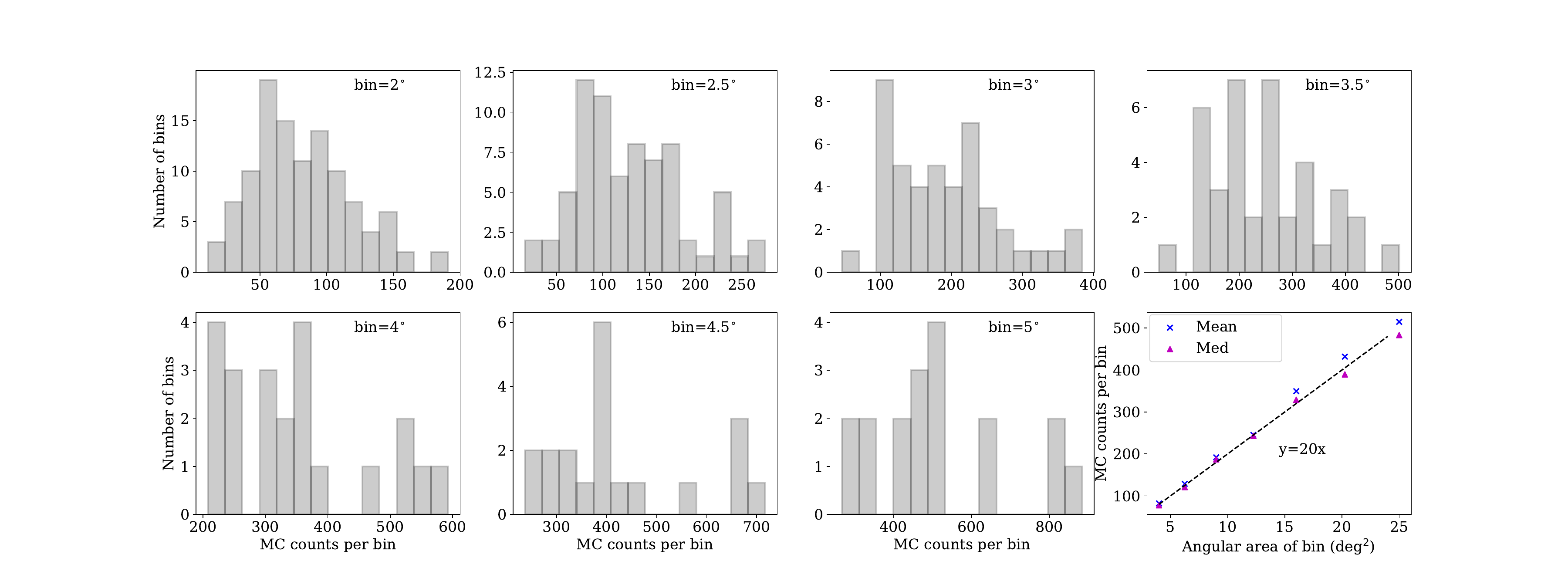}
    \caption{\textbf{Upper two panels}:The histogram distribution of the cloud-to-cloud velocity dispersions of $\sigma_{\rm bin}$ (gray) and $\sigma_{\rm bin,w}$ (green), 
    which is calculated at bin sizes of 2$^{\circ}$, 2.5$^{\circ}$, 3$^{\circ}$, 3.5$^{\circ}$, 4$^{\circ}$, 4.5$^{\circ}$ and 5$^{\circ}$, respectively, in the `Near' region.
    In the right-lower panel, the corresponding mean and median values of $\sigma_{\rm bin}$ (black) and $\sigma_{\rm bin,w}$ (green) at each bin size are presented. 
    \textbf{Lower two panels}:The histogram distribution of the MC number, 
    which is also counted within bin sizes of 2$^{\circ}$, 2.5$^{\circ}$, 3$^{\circ}$, 3.5$^{\circ}$, 4$^{\circ}$, 4.5$^{\circ}$ and 5$^{\circ}$, respectively, 
    in the `Near' region. In the right-lower panel, the corresponding mean and median values of MC number at each bin area are presented. \label{fig:cl_hist_near}}
\end{figure*}

Figure \ref{fig:cl_hist_near} presents histograms of $\sigma_{\rm bin}$ and $\sigma_{\rm bin, w}$, 
calculated for bin sizes of 2$^{\circ}$, 2.5$^{\circ}$, 3$^{\circ}$, 3.5$^{\circ}$, 4$^{\circ}$, 4.5$^{\circ}$ and 5$^{\circ}$. 
The mean and median values corresponding to each bin size are also illustrated. 
We find the mean and median values of $\sigma_{\rm bin}$ gradually increase from 7.0 km s$^{-1}$ to 7.7 km s$^{-1}$ 
as bin sizes increase from 2$^{\circ}$ to 4$^{\circ}$. 
However, these values plateau at around $\sim$ 7.6 km s$^{-1}$ for bin sizes of 4$^{\circ}$ -- 5$^{\circ}$. 
The $\sigma_{\rm bin, w}$ shows a similar trend, increasing from 5.6 km s$^{-1}$ to 6.5 km s$^{-1}$ as 
bin sizes progress from 2$^{\circ}$ to 4$^{\circ}$, stabilizing at approximately 6.2 km s$^{-1}$ for bin sizes from 4$^{\circ}$ to 5$^{\circ}$. 
Notably, the value for the 3.5$^{\circ}$ bins experiences a slight drop, 
which may be attributed to the arrangement of bins along the latitudes. 
Bins sized 3$^{\circ}$ cover the latitude range of -4.5$^{\circ}$ to 4.5$^{\circ}$,
whereas bins sized 3.5$^{\circ}$ cover the range of -5.25$^{\circ}$ to 5.25$^{\circ}$, 
and 4$^{\circ}$ bins cover a range of -4$^{\circ}$ to 4$^{\circ}$. 
As shown in Figure \ref{fig:cl_map_near}, the $\sigma_{\rm bin}$ value at higher latitudes 
tends to be lower. The changes in cloud-to-cloud velocity dispersions with respect to Galactic latitudes 
will be further discussed in Section \ref{sec:lat}.

Since systematic errors on the $\sigma_{\rm bin}$ and $\sigma_{\rm bin, w}$ are related to the number of clouds in each bin.
Figure \ref{fig:cl_hist_near} illustrates the number of extracted $^{12}$CO clouds counted in each bin, 
with sizes ranging from 2$^{\circ}$ to 5$^{\circ}$ in increments of 0.5$^{\circ}$. 
We find that the counts of MCs in the bins with anguar size of 2$^{\circ}$, 2.5$^{\circ}$, 3$^{\circ}$, and 3.5$^{\circ}$ 
show relatively concentrated distributions. 
In contrast, the distributions for the bins sized 4$^{\circ}$ and 5$^{\circ}$ are more dispersed, 
likely due to the limited number of bins available in these areas.
Additionally, the mean and median values increase almost linearly with a slope of 20, 
as the bin angular areas expand from 2$^{\circ}\times$2$^{\circ}$ to 5$^{\circ}\times$5$^{\circ}$.  
This suggests that the majority of clouds in the `Near' region are evenly distributed.

\subsection{Cloud-to-cloud velocity dispersions of MCs with different scales \label{sec:popus}}
In the dynamical evolution of MCs, 
cloud-to-cloud velocity dispersion and the clouds' size and spatial distributions 
determine merger rates between clouds. 
Conversely, mergers of MCs likely affect the cloud-to-cloud velocity dispersions, 
clouds' size and spatial distributions as well. 
If we model the merging process of MCs as inelastic collisions, 
it suggests that mergers between small-scale MCs result in the formation of large-scale MCs, 
accompanied by a loss of kinetic energy. 
In constrast, the destruction of these large-scale MCs, caused by the internal or external forces from dynamical activities 
such as stellar feedbacks, may lead to the creation of small-scale MCs with higher kinetic energy. 
Therefore, we will further investigate the cloud-to-cloud velocity dispersions among MCs with different scales.
 
\begin{figure*}[ht]
    \centering
    \includegraphics[width=1.0\textwidth, trim=50 40 30 90, clip]{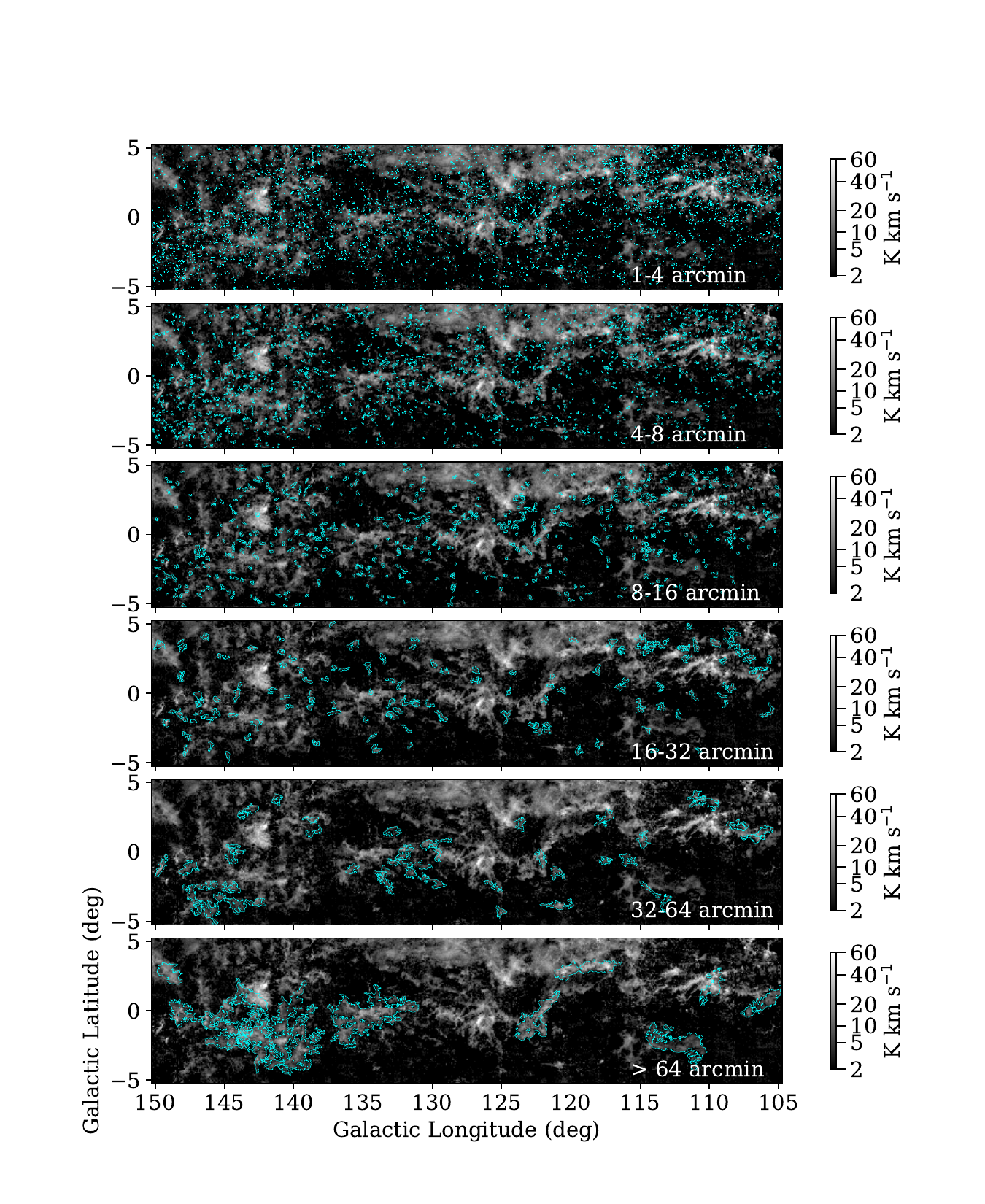}
    \caption{The spatial distribution of MCs in the `Near' region with angular sizes of 1 -- 4, 4 -- 8, 8 -- 16, 16 -- 32, 32 -- 64 arcmin, 
    and larger than 64 arcmin, respectively, are projected on the Galactic l-b plane. 
    Several zoomed-in images of MCs with angular sizes of 1 -- 4 arcmin, along with their averaged spectral lines, 
    are presented in the appendix of Figure \ref{fig:cloudS_example}. MCs with angular sizes of 1 -- 8 arcmin are classified as type $S$, 
    those having angular sizes of 8 -- 32 srcmin are classified as type $M$, and MCs greater than 32 arcmin are classified as type $L$.
    The background colormap is the velocity-integrated intensity of $^{12}$CO(1-0) line emission, 
    whose integrated velocity range is (-30 25) km s$^{-1}$.  
    It should be noted that an extracted $^{12}$CO structure across the region with l ranging from 105$^{\circ}$ and 145$^{\circ}$, 
    and a velocity range from -25 to 15 km s$^{-1}$, as illustrated in Figure 6 of \cite{Yan2021}, has been excluded from this research. 
    As a result, some emissions in the velocity-integrated intensity maps do not correspond to the plotted samples of MCs and are 
    associated with this excluded structure. \label{fig:cloud_popus}}
\end{figure*}

Figure \ref{fig:cloud_popus} presents the distributions of $^{12}$CO clouds in the longitude-latitude (l-b) plane, 
categorized by angular sizes, where l$_{n}$ = 2$^{n}$ - 2$^{n+1}$ arcmin for n=0,1,2,3,4,5,6, respectively. 
The angular size l=$\sqrt{A}$, where $A$ is the observed angular area of an extracted MC.  
We find clouds with angular sizes between 1 -- 4 arcmin and 4 -- 8 arcmin are evenly distributed in this region. 
However, for $^{12}$CO clouds with angular sizes of 8 -- 16 arcmin and 16 -- 32 arcmin,  
the distribution remains generally even but exhibits some concentration in specific local areas, 
such as around $l$ = 115$^{\circ}$ and 140$^{\circ}$. 
In contrast, the distribution of $^{12}$CO clouds greater than 32 arcmin is relatively uneven, 
showing signs of crowding and structured patterns. 
It is important to note that these large-scale clouds are sparse within this defined area 
and may only represent a localized distribution.  
Their distribution across a larger area will be further discussed in Section \ref{sec:lat}.

We categorize these MCs into three types: $S$, $M$, $L$, 
based on their sizes and their distributions on the Galactic plane.  
Type $S$ MCs have angular sizes ranging from 1 to 8 arcmin, 
totaling 8,792 clouds, which account for 91.4$\%$ of the total samples in the `Near' region. 
These clouds are almost uniformly distributed within the defined area. 
Type $M$ MCs have angular sizes from 8 to 32 arcmin, consisting of 766 clouds that make up about 8$\%$ of the total. 
These clouds are generally evenly distributed, there are some slight local unevenness in their distribution. 
Type $L$ MCs, with angular sizes greater than 32 arcmin, comprise only 59 clouds, representing the remaining 0.6$\%$ of the total. 
In the `Near' region, these clouds have a typical kinematic distance of $\sim$ 0.5 kpc. 
Consequently, the linear scale for 1 arcmin is about 0.15 pc. 
These parameters of MCs in Type $S$, $M$, and $L$ are also listed in Table \ref{tab:t_fac}. 
It is important to note that the linear scales for MCs are based on a canonical distance 
rather than the exact distance for each cloud. Additionally, the angular sizes of MCs are calculated from observed angular areas, 
which may be underestimated due to the projection effect.   

\begin{figure}[ht]
    \plotone{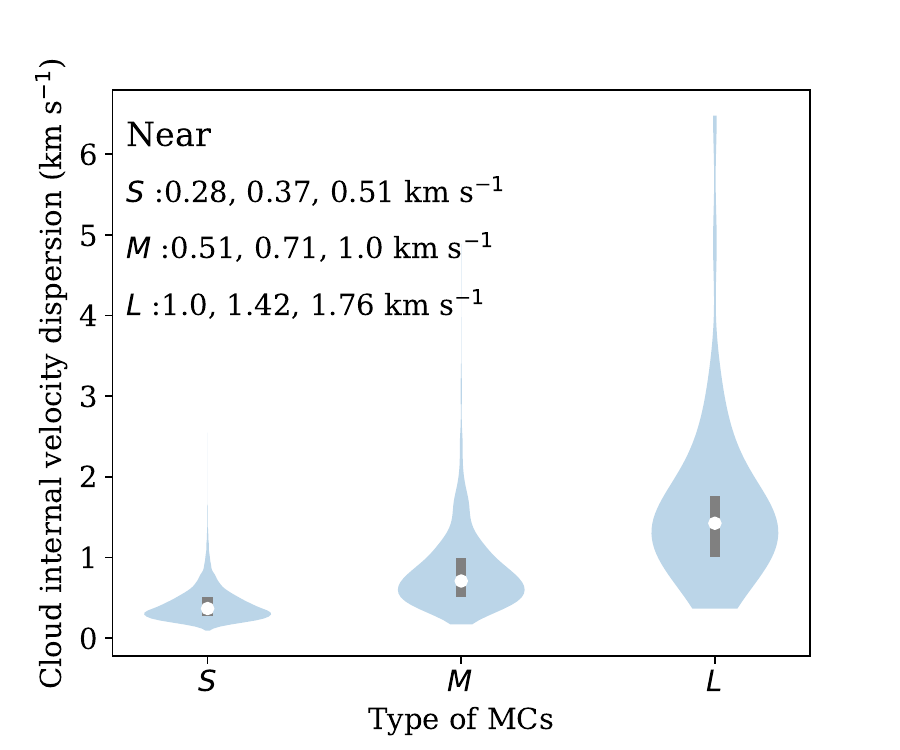}
    \caption{The distribution of the internal velocity dispersions ($\sigma_{\rm in}$) of MCs in Type $S$, $M$, $L$ of the `Near' region. 
    Among that white dots show median values and gray bars indicate the interquartile ranges. 
    The corresponding values of medians and quartiles at 25$\%$ and 75$\%$ are specifically noted in the upper-left corner. \label{fig:inner_vsigma_near}}
\end{figure}

Figure \ref{fig:inner_vsigma_near} displays the distributions of internal velocity dispersions $\sigma_{\rm in}$ for the MCs 
categorized into three types $S$, $M$, and $L$. 
The median values, along with quartiles at the 25th and 75th percentiles, are also provided. 
For type $S$ MCs, the median internal velocity dispersion is 0.37 km s$^{-1}$, 
with an interquartile range between 0.28 and 0.51 km s$^{-1}$. 
For type $M$ MCs, the median internal velocity dispersion is 0.71 km s$^{-1}$, 
with an interquartile range of 0.51 to 1.0 km s$^{-1}$. 
For type $L$ MCs, the median value is 1.35 km s$^{-1}$, 
with an interquartile range of 0.96 to 1.8 km s$^{-1}$. 
It should be noted that $^{12}$CO lines are usually optically thick, 
and this opacity can affect the measured internal velocity dispersions. 
\cite{Yuan2023b} have estimated the broadening factors due to the $^{12}$CO line opacity based on a sample of MCs, 
resulting in a median value is about 1.33. 
Our findings indicate that the internal velocity dispersions of type $M$ MCs are roughly double those of type $S$ MCs, 
and the values for type $L$ clouds are approximately twice those of type $M$ clouds. 
According to the power-law relationship between velocity dispersions ($\sigma$) and linear scales of MCs ($l$), 
as outlined by \cite{Larson1981} with $\sigma \approx l^{0.38}$ and by \cite{Solomon1987} with $\sigma \approx l^{0.5}$, 
the linear scales for type $M$ clouds are about four times larger than those for type $S$ clouds, as well as the linear scales for 
clouds in type $L$ compared to type $M$. 
The mean angular sizes for type $S$, $M$, and $L$ clouds are 3.6, 13.6, and 70.1 arcmin, respectively. 
In addition, the median values are 3.0, 10.8, and 46.1 arcmin, respectively. 
We find that the observed differences in angular sizes among type $S$, $M$, and $L$ are also about four times, 
which is overall consistent with those from linear scales estimated from their internal velocity dispersions.

\begin{figure*}[ht]
    \centering
    \includegraphics[width=1.0\textwidth, trim=120 30 120 30, clip]{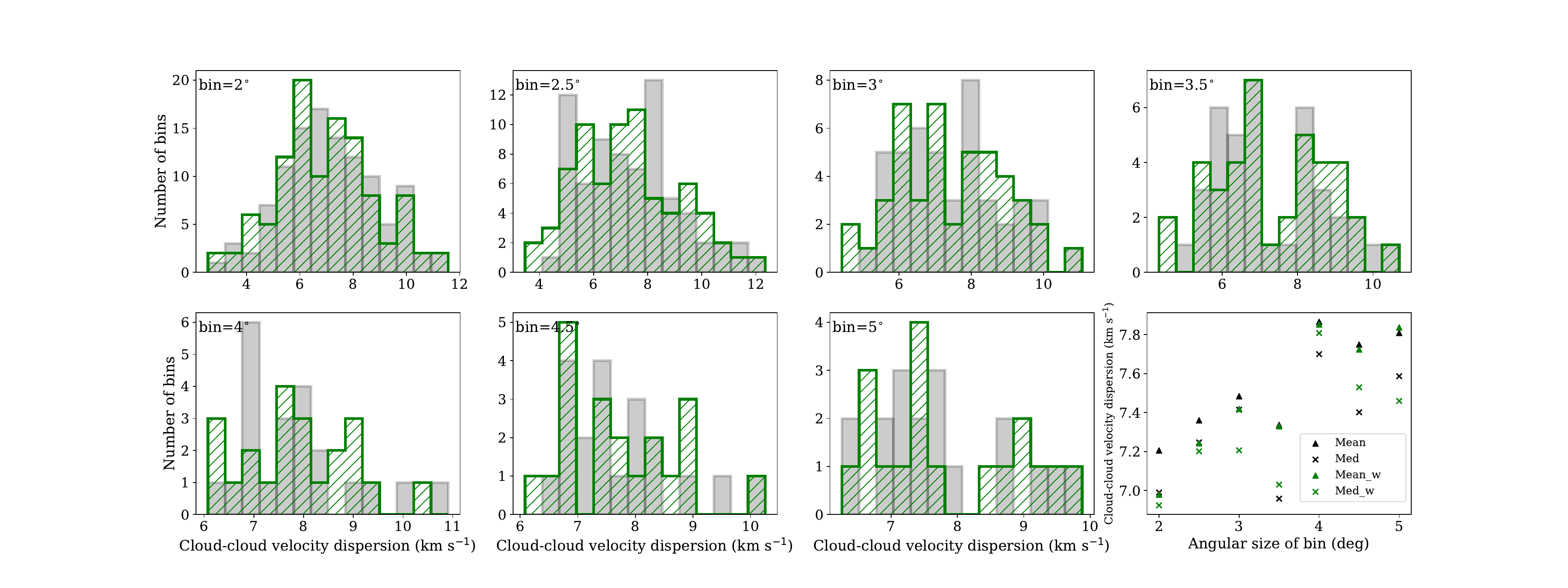}
    \includegraphics[width=1.0\textwidth, trim=120 30 120 30, clip]{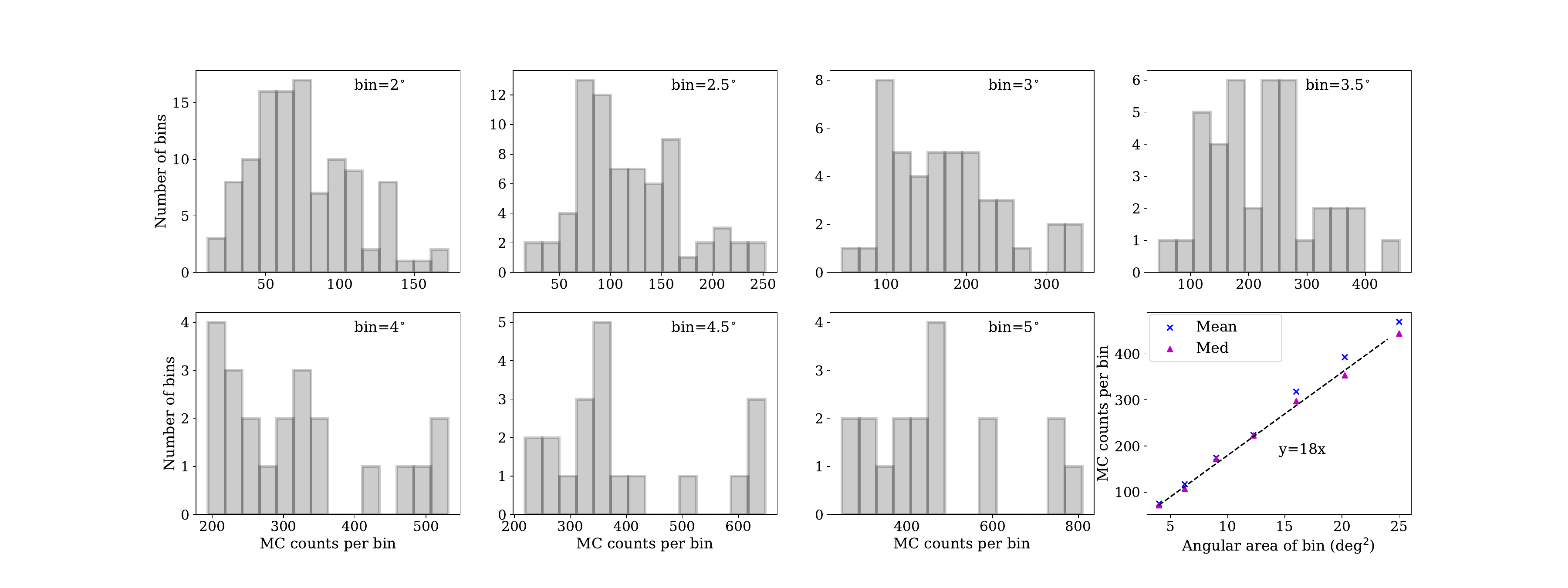}
    \caption{Same as Figure \ref{fig:cl_hist_near}, but for MCs with angular sizes ranging from 1 to 8 arcmin (Type $S$) in the `Near' region \label{fig:hist_popu1}}
\end{figure*}

After classifying MCs into types $S$, $M$, and $L$, 
we further calculated cloud-to-cloud velocity dispersions among MCs in these types.
For MCs classified as type $S$, Figure \ref{fig:hist_popu1} presents the distributions of their $\sigma_{\rm bin}$ and $\sigma_{\rm bin,w}$ values, 
calculated for bin sizes of 2$^{\circ}$, 2.5$^{\circ}$, 3$^{\circ}$, 3.5$^{\circ}$, 4$^{\circ}$, 4.5$^{\circ}$ and 
5$^{\circ}$. We find that $\sigma_{\rm bin}$ and $\sigma_{\rm bin,w}$ values for bins 
sized from 2$^{\circ}$ to 3.5$^{\circ}$ have relatively concentrated distributions. 
In contrast, the distributions for bin sizes of 4$^{\circ}$ -- 5$^{\circ}$ are somewhat dispersed, 
likely due to the limited number of bins in this defined area. 

In Figure \ref{fig:hist_popu1}, we also present the mean and median values of $\sigma_{\rm bin}$ and $\sigma_{\rm bin,w}$ for each bin size. 
We find the mean and median values of $\sigma_{\rm bin}$ increase from $\sim$ 7.2 km s$^{-1}$ to 7.8 km s$^{-1}$ as 
the bin size increases from 2$^{\circ}$ to 4$^{\circ}$, and they remain around 7.8 km s$^{-1}$ for bin sizes of 4$^{\circ}$ to 5$^{\circ}$. 
The mean and median values of $\sigma_{\rm bin, w}$ exhibit a similar trend, 
although they are generally about 0.2 km s$^{-1}$ lower. 
Due to the approximate scales of clouds in Type $S$, the differences between $\sigma_{\rm bin}$ and $\sigma_{\rm bin, w}$ 
are less pronounced compared to those in the entire sample ($\sim$ 1.5 km s$^{-1}$). 
Furthermore, the distribution of $\sigma_{\rm bin}$ for Type $S$ closely resembles that of the entire samples 
since clouds in Type $S$ constitute about 91.4$\%$ of the whole sample in the `Near' region. 
We also noted a slight drop in the mean and median values for the bin size of 3.5$^{\circ}$. 
This could be due to the same reason mentioned in \ref{sec:MC_whole},   
particularly the sparse clouds are distributed in the higher latitude, 
as illustrated by the clouds in the range of $\mid b \mid$ = 4$^{\circ}$ - 5$^{\circ}$, shown in Figure \ref{fig:cloud_popus}.

Figure \ref{fig:hist_popu1} displays the number of $^{12}$CO clouds classified as Type $S$, 
counted in each bin with angular sizes ranging from 2$^{\circ}$ to 5$^{\circ}$ in increments of 0.5$^{\circ}$. 
The cloud count distributions for varying bin sizes are similar to those of the entire sample. 
The mean and median values increase almost linearly, 
albeit at a slightly lower slope of 18 compared to a slope of 20 for the entire sample. 
The similar trend observed between Type $S$ clouds and the overall sample 
indicates that Type $S$ clouds represent the majority of clouds in the entire sample 
and are uniformly distributed across the area.

\begin{figure*}[th]
    \centering
    \includegraphics[width=1.0\textwidth, trim=120 30 120 30, clip]{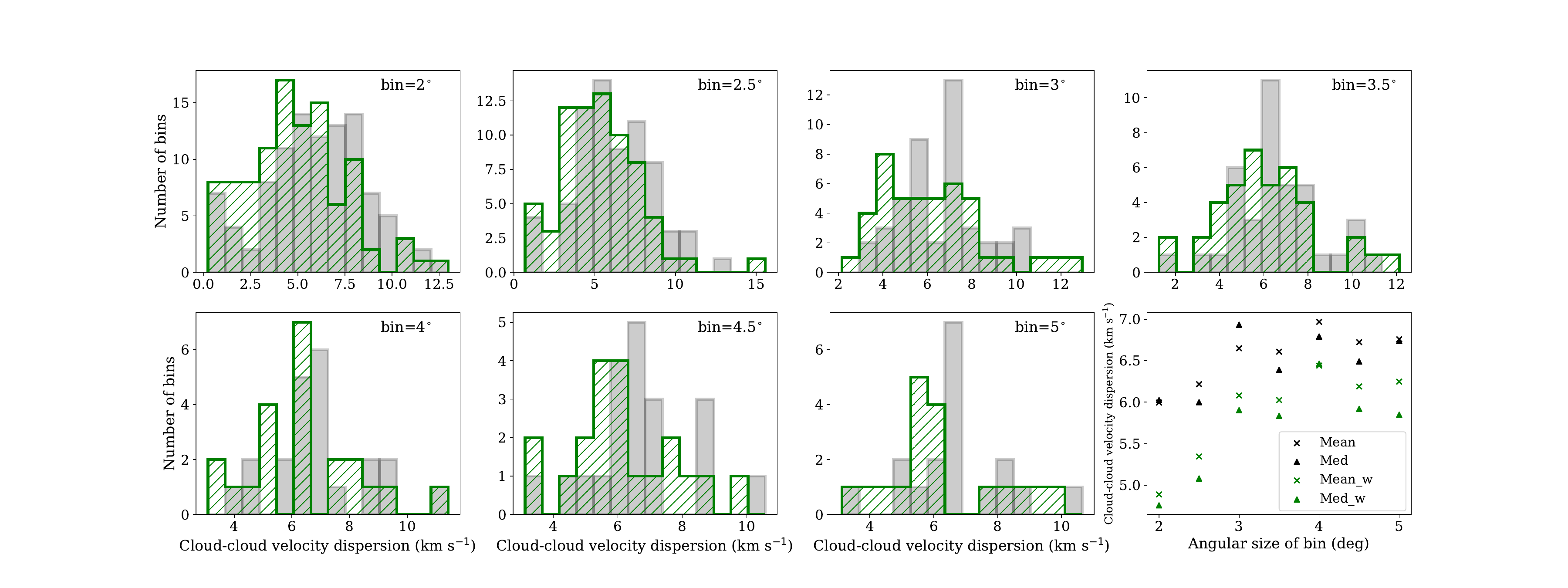} 
    \includegraphics[width=1.0\textwidth, trim=120 30 120 30, clip]{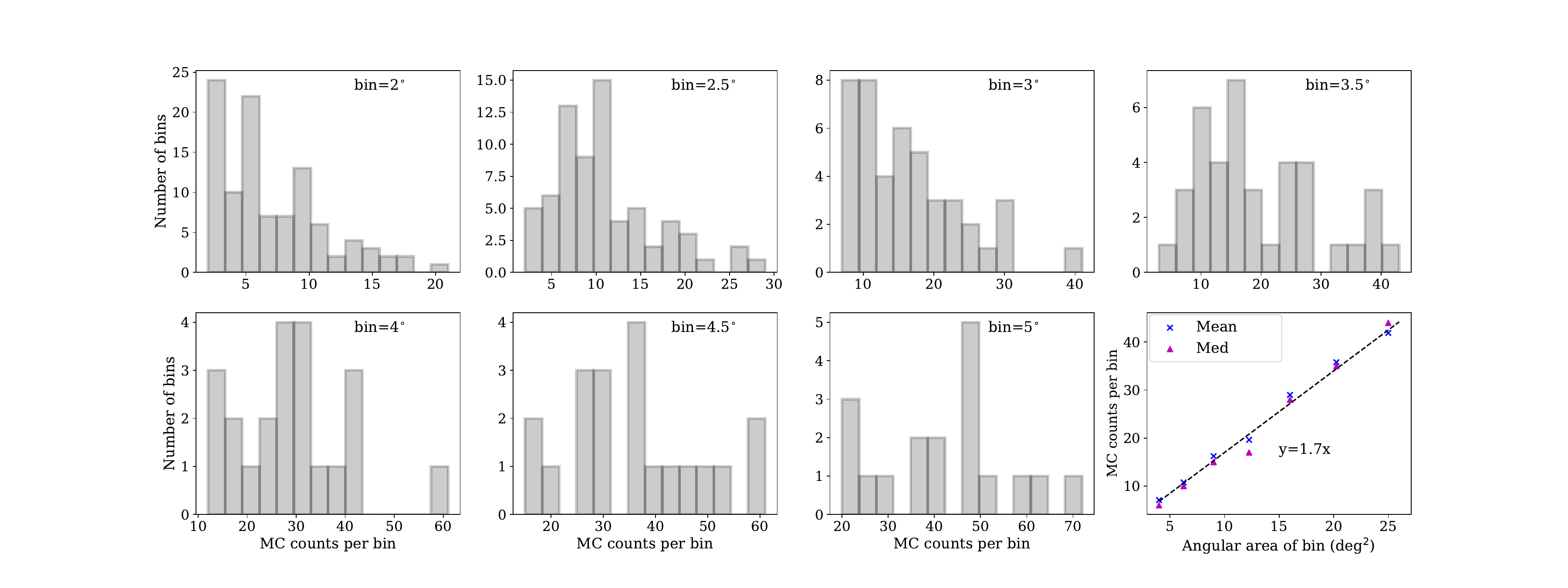}  
    \caption{Same with Figure \ref{fig:cl_hist_near}, but for MCs with angular sizes ranging from 8 to 32 arcmin (Type $M$) in the `Near' region. \label{fig:hist_popu2}}
\end{figure*}

For clouds in Type $M$, Figure \ref{fig:hist_popu2} presents the distributions of their $\sigma_{\rm bin}$ and $\sigma_{\rm bin,w}$ values, 
calculated for bin sizes of 2$^{\circ}$, 2.5$^{\circ}$, 3$^{\circ}$, 3.5$^{\circ}$, 4$^{\circ}$, 4.5$^{\circ}$ and 5$^{\circ}$. 
The mean and median values for each bin size are displayed as well. 
We found that the mean and median values of $\sigma_{\rm bin}$ increase from $\sim$ 6.0 km s$^{-1}$ to 6.7 km s$^{-1}$ 
as the bin size increases from 2$^{\circ}$ to 3$^{\circ}$, then plateau at around 6.7 km s$^{-1}$ for bin sizes from 3$^{\circ}$ to 5$^{\circ}$. 
The mean and median values of $\sigma_{\rm bin, w}$ follow a similar trend, but are generally lower by about 0.7 km s$^{-1}$.  

Figure \ref{fig:hist_popu2} also illustrates the number of the Type $M$ clouds counted in the specific bins. 
The overall number of Type $M$ clouds within this defined area is relatively small, 
constituting only about 8$\%$ of the entire sample.
For bins sized 2$^{\circ}$, the mean and median number of clouds per bin is about 6. 
These values rises to about 10 for bins sized 2.5$^{\circ}$.
Although the calculated values of $\sigma_{\rm bin}$ and $\sigma_{\rm bin, w}$ are initially relatively lower, 
they increase sharply, which may be influenced by statistical errors. 
For bin sizes ranging from 3$^{\circ}$ to 5$^{\circ}$, 
both mean and median values of the cloud number per bin rise from about 15 to 45.
The corresponding $\sigma_{\rm bin}$ remains steady at $\sim$ 6.7 km s$^{-1}$, 
while $\sigma_{\rm bin, w}$ remains around $\sim$ 6.0 km s$^{-1}$. 
For Type $L$ clouds, their number is insufficient to calculate their velocity dispersions in this defined region.
Therefore, we could not measure their $\sigma_{\rm bin}$ and $\sigma_{\rm bin, w}$ values in this study. 
A larger sample covering a wider area is necessary for accurate measurement. 

Overall, from the mean and median of $\sigma_{\rm bin}$ and $\sigma_{\rm bin, w}$ values across different bin sizes, 
we find that $\sigma_{\rm bin}$ values for Type $S$ clouds are systematically greater by about 0.9 km s$^{-1}$ compared to 
those for Type $M$ clouds. Additionally, the $\sigma_{\rm bin, w}$ values for Type $S$ clouds exceed 
those of Type $M$ by $\sim$ 1.4 km s$^{-1}$. 
This indicates that velocity dispersions among MCs with smaller scales tend to be higher than those at larger scales. 
This trend is consistent with the finding that $\sigma_{\rm bin}$ values for the entire samples 
are systematically greater than their corresponding $\sigma_{\rm bin, w}$ values, 
with large-scale clouds receiving higher weights. 

\begin{deluxetable*}{lccc}
    \tablecaption{MC samples in the Type $S$, $M$, and $L$. \label{tab:t_fac}}
    \tablewidth{0pt}
    \tablehead{\colhead{} & \colhead{Type $S$} & \colhead{Type $M$} & \colhead{Type $L$} \\}
    \startdata
    Number & 8792 (91.4$\%$) & 766 (8$\%$) & 59 (0.6$\%$) \\
    Angular size (arcmin) & 1 -- 8 & 8 -- 32 & $>$ 32 \\
    Median angular size (arcmin) & 3.0 & 10.8 & 46.1 \\
    Mean angular size (arcmin) & 3.6 & 13.6 & 70.1 \\
    Linear scale (pc) & 0.15 -- 1.2 & 1.2 -- 4.8 & $\gtrsim$ 4.8 \\
    Integrated flux (K km s$^{-1}$ arcmin$^{2}$) & 1.5$\times$10$^{5}$ & 3.7$\times$10$^{5}$ & 1.5$\times$10$^{6}$ \\
    Median $\sigma_{\rm in}$ (km s$^{-1}$)& 0.37 & 0.71 & 1.35 \\
    $\sigma_{\rm bin}$ (km s$^{-1}$) & 7.6$\pm$0.3 & 6.7$\pm$0.5 & $\text{- -}$  \\
    $\sigma_{\rm bin,w}$ (km s$^{-1}$) & 7.4$\pm$0.3 & 6.0$\pm$0.5 & $\text{- -}$  \\
    t$_{\rm merger}$ (Myr) & 0.3 -- 0.9 & 1 -- 2.5 & $\text{- -}$  \\
    t$_{\rm cross}$ (Myr) & $\sim$1 & $\sim$2 & $\gtrsim$ 5.0 \\
    \enddata
\tablecomments{The parameters for MCs in Type $S$, $M$, and $L$, including their number, angular sizes, linear scales, 
and total integrated fluxes, cloud-to-cloud velocity dispersion ($\sigma_{\rm bin}$ and $\sigma_{\rm bin,w}$), 
internal velocity dispersions ($\sigma_{\rm in}$), the estimated merger timescale (t$_{\rm merger}$) and cross timescale (t$_{\rm cross}$).}
\end{deluxetable*}

\section{Discussion}
\subsection{Expanding cloud-to-cloud velocity dispersions from the `Near' to `Far' Galactic segments \label{sec:lat}}
The results presented above focus on a defined segment, 
i.e. the `Near' region, where the Galactic longitude $l$ ranges from 104$^{\circ}$.75 
to 150$^{\circ}$.25, Galactic latitude $\mid b \mid <$ 5$^{\circ}$.25, 
and the line-of-sight velocity ($V_{\rm LSR}$) is between -30 km s$^{-1}$ and 25 km s$^{-1}$. 
This region is mainly located in the local arm and corresponds to a typical kinematic distance of $\sim$ 0.5 kpc.
In this area, we have found that the cloud-to-cloud velocity dispersions among smaller-scale MCs are greater than 
those among larger-scale MCs. 
However, we question whether this trend persists in other environments? 

\begin{figure*}[ht]
    \centering
    \includegraphics[width=1.0\textwidth, , trim=50 40 30 50, clip]{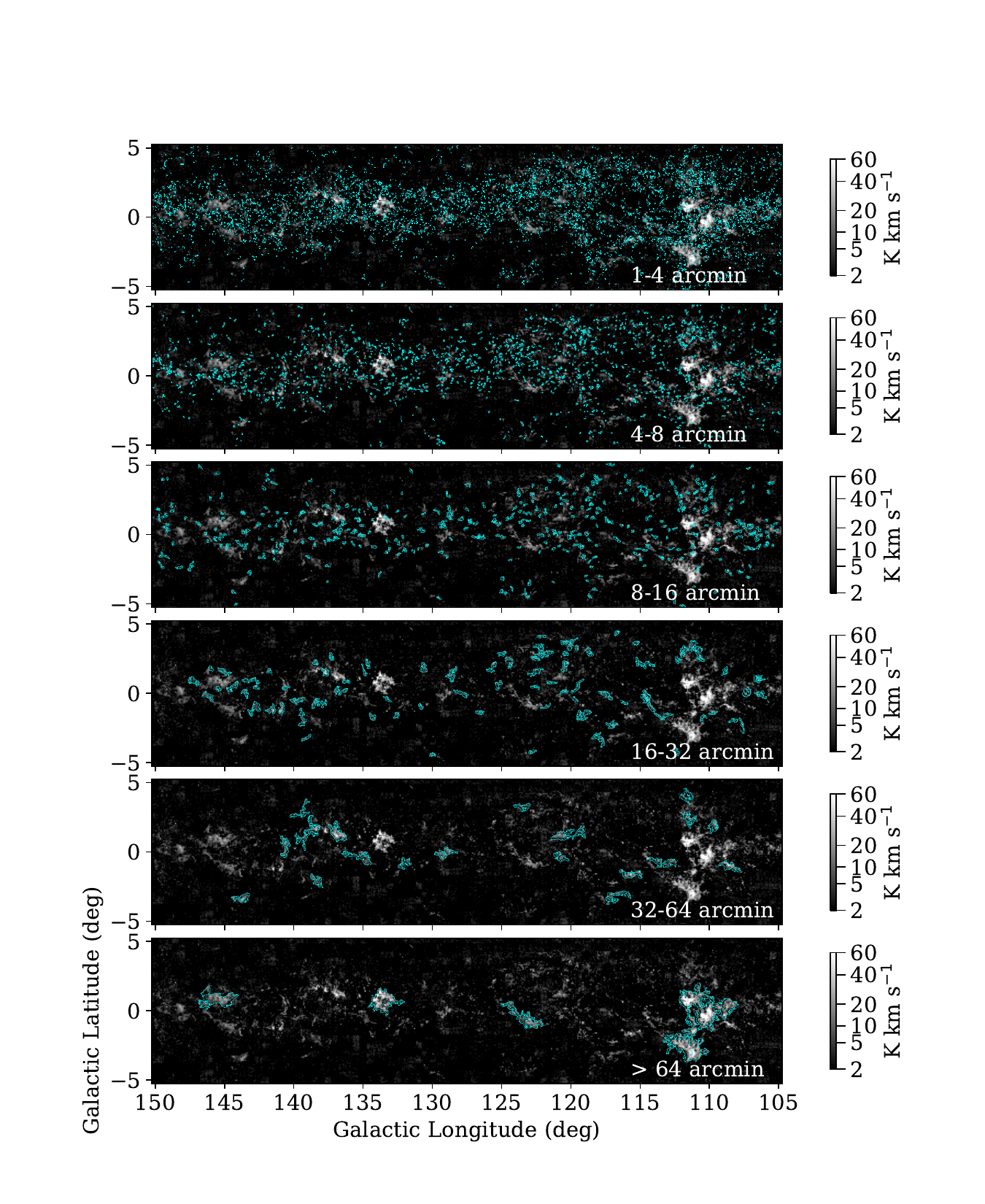} 
    \caption{The spatial distribution of MCs with angular sizes of 1 -- 4, 4 -- 8, 8 -- 16, 16 -- 32, 32 -- 64 arcmin, 
    and larger than 64 arcmin, respectively, in the `Far' region, are projected on the Galactic l-b plane. 
    MCs with angular sizes of 1 -- 8 arcmin are classified as type $S^{\prime}$, 
    those having angular sizes of 8 -- 32 srcmin are classified as type $M^{\prime}$, 
    and MCs greater than 32 arcmin are classified as type $L^{\prime}$. 
    The background colormap is the corresponding velocity-integrated intensity of $^{12}$CO(1-0) line emission, 
    whose integrated velocity range is (-95 -30) km s$^{-1}$. \label{fig:cloud_popus_far}}
\end{figure*}

To investigate this, we analyze the $^{12}$CO emission in the `Far' region, 
where l and b fall within the same range, but velocities are instead between -95 km s$^{-1}$ and -30 km s$^{-1}$. 
This region corresponds to a segment of the Perseus arm and has a typical kinematical distance of $\sim$ 2 kpc. 
Due to its greater distance, the `Far' region encompasses a larger physical area.  
Figure \ref{fig:cloud_popus_far} displays the distribution of the extracted $^{12}$CO clouds of the `Far' region, 
whose angular sizes are within 2$^{n}$ -- 2$^{n+1}$ arcmin, where n=0,1,2,3,4,5,6, respectively. 
On a larger linear scale, it is evident that the vertical distribution of MCs sharply decreases 
at the higher latitude, most MCs are concentrated within $\mid b \mid \lesssim$ 3 deg 
and are relatively sparse within the range of 3 $\lesssim \mid b \mid \lesssim$ 5 deg.  

\begin{figure}[ht]
    \plotone{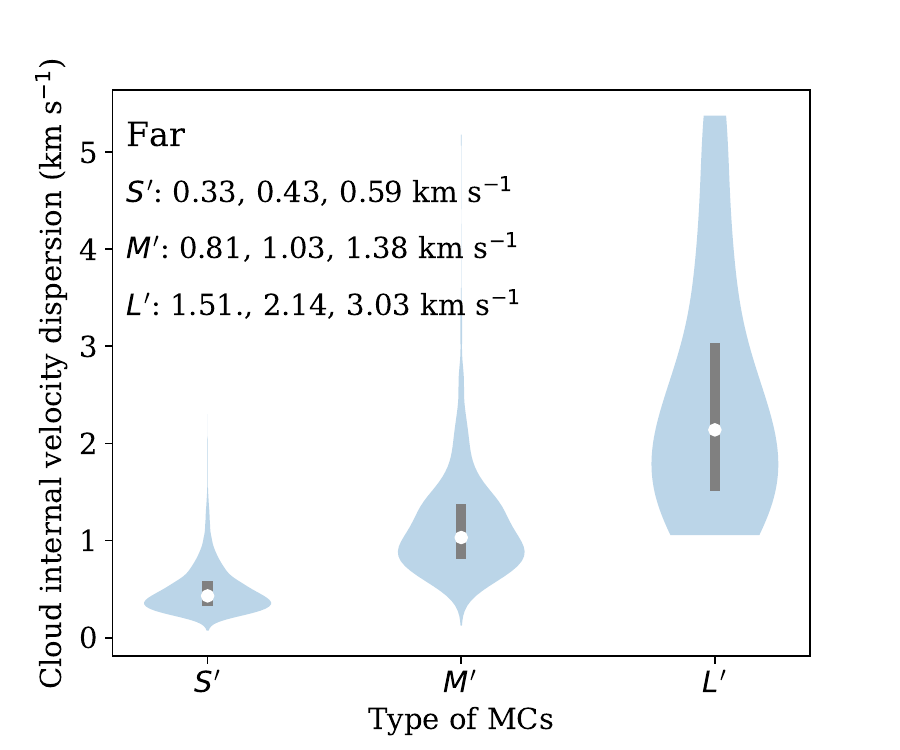}
    \caption{The distribution of the internal velocity dispersions ($\sigma_{\rm in}$) of MCs in Type $S^{\prime}$, $M^{\prime}$, $L^{\prime}$ of the `Far' region. 
    Among that the white dots show median values and gray bars indicate the interquartile ranges. 
    The corresponding values of the medians and quartiles at 25$\%$ and 75$\%$ are specifically noted in the upper-left corner. \label{fig:inner_vsigma}}
\end{figure}

Similar to the MCs in the `Near' region, we classify the extracted MCs in the `Far' region into three types: 
Type $S^{\prime}$ for MCs with angular sizes between 1 and 8 arcmin, 
Type $M^{\prime}$ for angular sizes between 8 and 32 arcmin, 
and Type $L^{\prime}$ for sizes larger than 32 arcmin.  
Out of the total sample of 8684 MCs in the `Far' region, 
8145 (93.8$\%$) are classified as Type $S^{\prime}$, 
514 (5.9$\%$) as Type $M^{\prime}$, and 25 (0.3$\%$) as Type $L^{\prime}$. 
In Figure \ref{fig:inner_vsigma}, we present distributions of internal velocity dispersions of MCs in each type, 
including the corresponding median values and quartiles at 25th and 75th percentiles. 
For MCs in Type $S^{\prime}$, the median internal velocity dispersion is 0.43 km s$^{-1}$, 
with an interquartile range between 0.33 and 0.59 km s$^{-1}$. 
For MCs in Type $M^{\prime}$, the median velocity dispersion is 1.03 km s$^{-1}$, 
with an interquartile range from 0.81 to 1.38 km s$^{-1}$. 
For MCs in Type $L^{\prime}$, the median value is 2.14 km s$^{-1}$, 
with an interquartile range from 1.51 to 3.03 km s$^{-1}$. 

We further calculate the cloud-to-cloud velocity dispersion among MCs of different types in the `Far' region, 
focusing on bins confined within $\mid b \mid \lesssim$ 3 deg, where most MCs are concentrated. 
Given that one arcmin corresponds to $\sim$ 0.6 pc in the `Far' region, 
we begin with a bin size of 1$^{\circ}$ and increase it to 3$^{\circ}$ by increments of 0.5$^{\circ}$, 
ensuring that their linear sizes are comparable to those in the `Near' region, 
meanwhile containing an adequate number of MCs to statistically analyze the cloud-to-cloud velocity dispersion. 

\begin{figure*}[ht]
    \centering
    \includegraphics[width=1.0\textwidth, trim=50 40 50 40, clip]{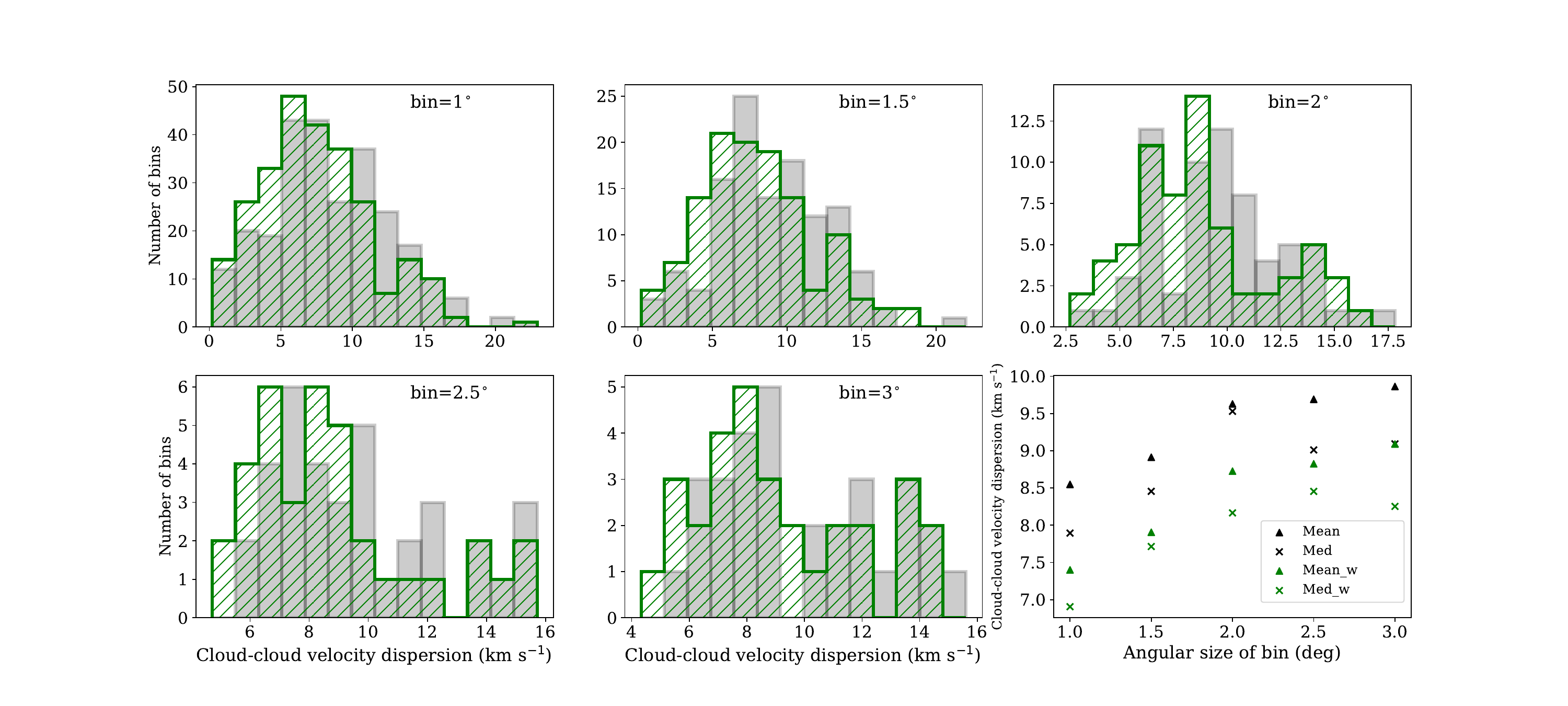}
    \includegraphics[width=1.0\textwidth, trim=50 40 50 40, clip]{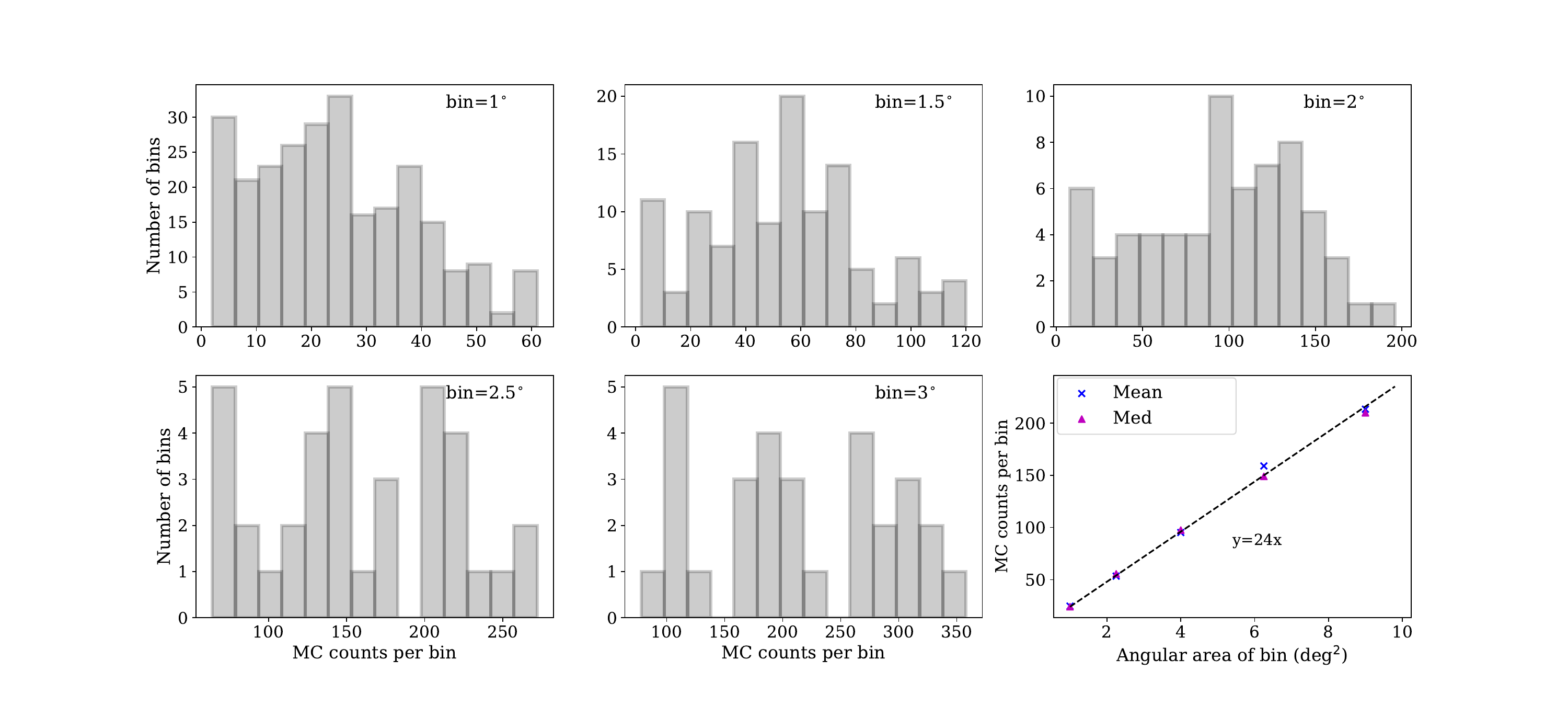}
    \caption{\textbf{Upper two panel}: the histogram distributions of cloud-to-cloud velocity dispersions of $\sigma_{\rm bin}$ 
    (gray) and $\sigma_{\rm bin,w}$ (green) for clouds with angular sizes of 1 -- 8 arcmin (\textbf{Type $S^{\prime}$}) in the `Far' region, 
    which is calculated at bin sizes of 1$^{\circ}$, 1.5$^{\circ}$, 2$^{\circ}$, 2.5$^{\circ}$, 3$^{\circ}$, respectively. 
    In the right-lower panel, the corresponding mean and median values of $\sigma_{\rm bin}$ and $\sigma_{\rm bin,w}$ at each bin size are presented. 
    \textbf{Lower two panel}: the histogram distribution of the number of MCs with angular sizes of 1 -- 8 arcmin (\textbf{Type $S^{\prime}$}) in the `Far' region, 
    which is counted within bin sizes of 1$^{\circ}$, 1.5$^{\circ}$, 2$^{\circ}$, 2.5$^{\circ}$, 3$^{\circ}$, respectively. 
    In the right-lower panel, the corresponding mean and median values of MC number at each bin area are presented. \label{fig:hist_popu1_far}}

\end{figure*}

For MCs in Type $S^{\prime}$, 
Figure \ref{fig:hist_popu1_far} displays the distributions of $\sigma_{\rm bin}$ and $\sigma_{\rm bin,w}$ values 
calculated with bin sizes of 1$^{\circ}$, 1.5$^{\circ}$, 2$^{\circ}$, 2.5$^{\circ}$, and 3$^{\circ}$, respectively.  
We find that the mean value of $\sigma_{\rm bin}$ increases from $\sim$ 8.5 to 9.5 km s$^{-1}$, 
as the bin sizes enlarges from 1$^{\circ}$ to 2$^{\circ}$, followed by a slower increase 
from $\sim$ 9.5 to 9.8 km s$^{-1}$ for bin sizes from 2$^{\circ}$ to 3$^{\circ}$. 
The median value of $\sigma_{\rm bin}$ shows a similar trend, 
albeit systematically lower by about 0.5 km s$^{-1}$. 
For both mean and median values of $\sigma_{\rm bin, w}$, similar trends are observed, 
but overall values are approximately 1 km s$^{-1}$ lower. 
Figure \ref{fig:hist_popu1_far} illustrates the number of MCs in each bin; 
for bins with an angular size of 1$^{\circ}$, 
the mean and median values of MCs' number within them are about 25. 
However, around 18$\%$ of the bins contain fewer than 10 MCs, 
resulting in greater statistic errors for $\sigma_{\rm bin}$ in these bins. 
The count of MCs for bins of sizes 1.5$^{\circ}$ and 2$^{\circ}$ shows more concentrated distributions, 
while bins of 2.5$^{\circ}$ and 3$^{\circ}$ exhibit more dispersed distributions, 
likely due to the limited number of bins in this area.
The mean and median values almost linearly increase with 
a slope of 24 as bin areas increase from 1$^{\circ}\times$1$^{\circ}$ to 3$^{\circ}\times$3$^{\circ}$. 
   
\begin{figure*}[ht]
    \centering
    \includegraphics[width=1.0\textwidth, trim=50 40 50 40, clip]{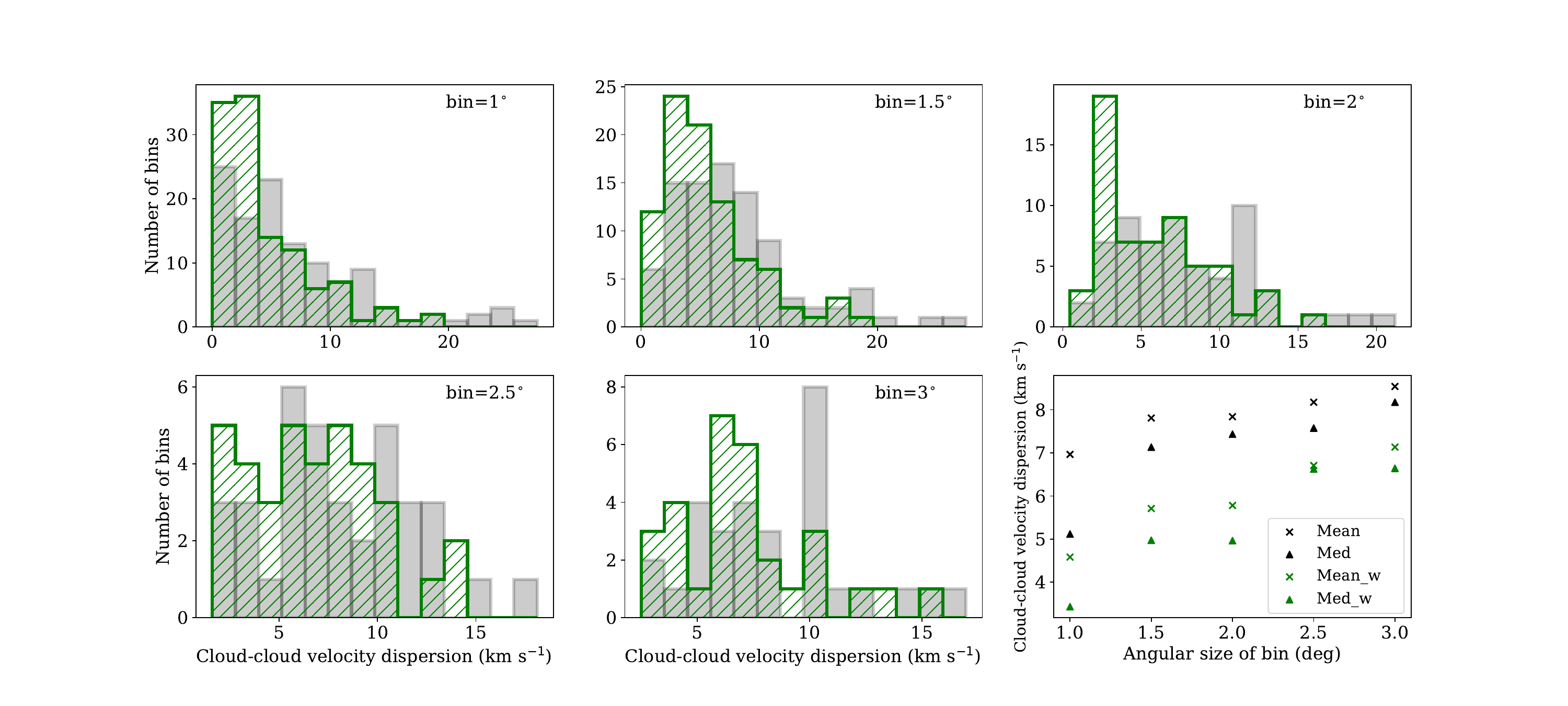}
    \includegraphics[width=1.0\textwidth, trim=50 40 50 40, clip]{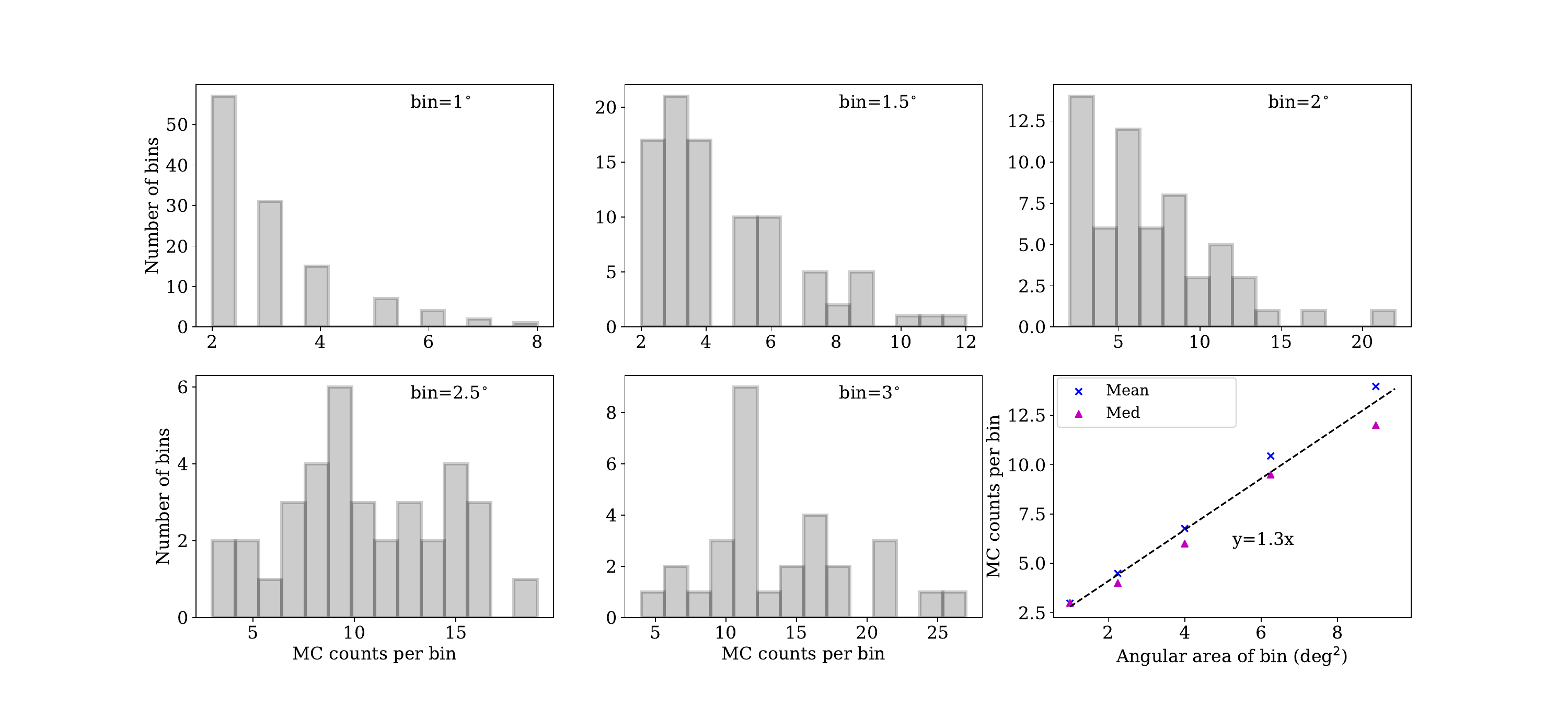}  
    \caption{Same with Figure \ref{fig:hist_popu1_far}, but for MCs with angular sizes ranging from 8 to 32 arcmin (\textbf{Type $M^{\prime}$}) in the `Far' region. \label{fig:hist_popu2_far}}
\end{figure*}

For MCs in Type $M^{\prime}$,
Figure \ref{fig:hist_popu2_far} presents distributions of their $\sigma_{\rm bin}$ and $\sigma_{\rm bin,w}$ 
calculated for bin sizes of 1$^{\circ}$, 1.5$^{\circ}$, 2$^{\circ}$, 2.5$^{\circ}$, and 3$^{\circ}$. 
The mean and median values of $\sigma_{\rm bin}$ increase from $\sim$ 7 to 8 km s$^{-1}$, 
as bin sizes increase from 1$^{\circ}$ to 3$^{\circ}$. 
Meanwhile, the mean and median values of $\sigma_{\rm bin, w}$ gradually increase from 5 to 6.2 km s$^{-1}$, 
as bin sizes increase from 1.5$^{\circ}$ to 3$^{\circ}$. 
The $\sigma_{\rm bin}$ and $\sigma_{\rm bin, w}$ exhibit similar trends, 
but $\sigma_{\rm bin, w}$ values are generally about 2 km s$^{-1}$ lower. 
The number of MC in Type $M^{\prime}$ comprises only about 5.9$\%$ of the whole `Far' samples.
Figure \ref{fig:hist_popu2_far} presents the number of MCs in each bin, 
showcasing that many bins contain fewer than 10 MCs, 
particularly for bins with angular areas of 1$^{\circ}\times$1$^{\circ}$ -- 2$^{\circ}\times$2$^{\circ}$. 
This results in greater statistical errors for $\sigma_{\rm bin}$ and $\sigma_{\rm bin, w}$ 
due to the limited counts of MCs in each bin. 

Compared with MCs in Type $S^{\prime}$ and $M^{\prime}$ of the `Far' region, 
we find that $\sigma_{\rm bin}$ values for Type $S^{\prime}$ are systematically greater by about 1.5 km s$^{-1}$ 
than those for clouds in Type $M^{\prime}$. 
Additionally, the $\sigma_{\rm bin, w}$ values in Type $S^{\prime}$ are 
greater by $\sim$ 2 km s$^{-1}$ compared to those in Type $M^{\prime}$. 
That further indicates that velocity dispersions between molecular clouds among smaller scales have a higher value, 
compared with clouds with larger scales.

Additionally, we find the measured cloud-to-cloud velocity dispersions in the `Far' regions are 
greater than those in the `Near' region. There are several influencing factors. 
One significant issue is that gas environments in the Local and Perseus arms likely differ \citep{Du2017, Sun2020}. 
The Perseus arm is considered one of the dominant spiral arms of the Milky Way, 
containing a large number of massive star-forming regions in the second quadrant \citep{Reid2019, Xu2023}. 
In contrast, the Local arm appears to be an isolated arm segment \citep{Xu2016, Reid2019, Xu2023}.
The typical kinematic distance is approximately 0.5 kpc for the 'Near' region 
and about 2 kpc for the 'Far' region. 
Due to the beam dilution effect, the extracted MC samples in different distances could be influenced. 
In the `Far' region, smaller clouds may not be detected because they have relatively low line intensities 
or they may be blended together with other smaller clouds. 
Consequently, the linear sizes of extracted Type $S^{\prime}$ MCs in the `Far' region 
cannot be assumed to be simply four times those of Type $S$ MCs in the `Near' region. 
The observational limitation on the MC samples are further discussed in the appendix of Section A. 
Additionally, there are considerable differences in spatial sizes of the bins with the same angular size 
between the `Near' and `Far' regions. 
This distance difference leads to an increase by a factor of 16 for the projected areas and 
a factor of 64 for the spatial volumes in the 'Far' region. 
As a result, the cloud-to-cloud velocity dispersions estimated from larger-size bins are 
likely more influenced by the systematic motions of clouds and may be overestimated. 
Despite these complexities, it remains a consistent finding that the velocity dispersion among smaller-scale MCs 
is generally higher than that among larger-scale MCs, regardless of whether they are in the 'Near' or 'Far' region.

\subsection{The estimated merger rate of molecular clouds}

As we proposed an alternative picture for the assembly and destruction of MCs: 
the transient process of MCs occurs through slow mergers or splits among the `fundamental blocks'.
The merger rate of individual MCs, which reflects the frequency of interactions among these clouds, 
is crucial for quantitatively characterizing this process. 
Based on the measured cloud-to-cloud velocity dispersion and the spatial distribution of MCs, 
we estimate the merge rate between these clouds. 
We also compare the merge timescale ($t_{\rm merge}$) of MCs with 
the internal crossing timescale ($t_{\rm cross}$) within MCs.

The merger rate depends on cloud-to-cloud velocity dispersion ($\sigma_{\rm cl}$), 
and the number and size distributions of MCs. 
The merger timescale can be expressed as: $t_{\rm merge} \propto \lambda/\sigma_{cl}$, 
where $\lambda$ represents the typical inter-cloud separation. 
In our observations, MCs categorized as Types $S$ and $M$ are evenly distributed across the study area, 
whose number account for 99.4$\%$ of the entire sample.  
According to that, we treat these MCs as point particles to roughly estimate the separations between them. 
Assuming a uniform distribution of MCs in a three-dimensional(3D) spatial space, 
we can define $\lambda_{\rm 3D} = (V/N)^{1/3}$, where $V$ is the volume of the defined segment and 
$N$ is the whole number of MCs in this segment. 
For the `Near' segment, the typical kinematic distance is about 500 pc, 
therefore, 1 arcmin corresponds to $\sim$ 0.15 pc. 
The latitude ranges from - 5.25 to 5.25 deg, yielding a linear scale of about 95 pc, 
which is nearly equivalent to the thickness of the Milky Way's thin disk \citep{Su2021}. 
We also assume that the arm segment width is around 95 pc. 
The longitude spans from 105 deg to 150 deg, resulting in a linear scale of about 400 pc. 
Overall, we identified a total of 9,617 $^{12}$CO MCs in this `Near' region.  
Consequently, the estimated value of $\lambda_{\rm 3D}$ is about 7 pc. 
However, we should consider that the MCs tend to concentrate on several layers. 
Under this assumption, the projected 2D separation ($\lambda_{\rm 2D}$) between MCs can be represented as 
the minimal separation of these MCs, given by $\lambda_{\rm 2D} = (A/N)^{1/2}$, 
where $A$ is the projected area of a segment on the l-b plane. 
Hence, the estimated value of $\lambda_{\rm 2D}$ is about 2 pc. 
Overall, the spatial separations between MCs are around 2 -- 7 pc. 
We should note that the above values mostly represent the separations among Type $S$ MCs, 
whose number accounts for 91.4$\%$ of the whole sample. 
For MCs in Type $S$, the value of $\sigma_{\rm cl}$ is about 7.5 km s$^{-1}$, 
leading to an estimated $t_{\rm merge}$ of around 0.3 -- 0.9 Myr.
For 761 MCs in Type $M$, the separations among them are about 7 -- 17 pc, 
with their $\sigma_{\rm cl}$ is about 6.7 km s$^{-1}$, 
resulting in a $t_{\rm merge}$ of about 1 -- 2.5 Myr.
Consider their scales of 1.2 -- 4.8 pc corresponding to the collision cross-sections, 
this $t_{\rm merge}$ value is likely overestimated by a factor of $\sim$ 0.2.
However, both Type $S$ and $M$ MCs are evenly distributed across the region, 
interactions between Type $S$ and $M$ MCs are inevitable and even more probable. 
Considering their separations and cross-sections, 
the $t_{\rm merge}$ between Type $S$ and $M$ MCs, as well as Type $S$ and $L$ MCs, 
is less than 0.3 -- 0.9 Myr. 
Overall, we find that the $t_{\rm merge}$ between Type $S$ MCs is shorter than that between Type $M$ MCs. 
Additionally, the $t_{\rm merge}$ between Type $S$ and $M$ MCs, along with Type $S$ and $L$ MCs, 
is shorter than that for Type $S$ MCs alone. 
Overall, this indicates that the interaction timescale for Type S-S, Type S-M, and Type S-L is less than $\sim$ 1 Myr.

For the crossing time for an MC, we can express it as: 
$t_{\rm cross} \propto l/\sigma_{\rm in}$, where $l$ represents the scale of the MC, 
$\sigma_{\rm in}$ is the internal velocity dispersion of the MC.
The median internal velocity dispersions of MCs in Type $S$, $M$, and $L$ 
are approximately 0.37, 0.71, and 1.35 km s$^{-1}$, respectively. 
Their median linear scales are roughly 0.45, 1.5, and 6.9 pc, respectively. 
Based on these values, we find that $t_{\rm cross}$ is $\sim$ 1 Myr for Type $S$ MCs 
and $\sim$ 2 Myr for Type $M$ MCs, while it exceeds $\sim$ 5 Myr for Type $L$ MCs. 
For the majority of MCs, such as MCs in Type $S$ and $M$, 
the crossing timescale is comparable to or even greater than the merger timescale.  
Notably, for Type $L$ MCs, the crossing time is several times greater than $t_{\rm merge}$. 
This suggests that MCs are most likely to be transient, with the constituent gas of individual MCs 
changing over short timescales of less than $\sim$ 1 Myr due to mutual interactions with the surrounding MCs.
Large-scale MCs, in particular, tend to undergo multiple mergers. 
We have also observed that 40$\%$ of MCs with double and triple $^{13}$CO structures 
display distinct velocity discontinuities \citep{Yuan2024}. 
Furthermore, the relative velocities measured along the line of sight between $^{13}$CO structures within MCs 
are generally less than $\sim$ 5 km s$^{-1}$. 
This maximal value is lower than the measured $\sigma_{\rm cl}$, 
which are approximately $\sim$ 7.6 km s$^{-1}$ for Type $S$ and $\sim$ 6.7 km s$^{-1}$ for Type $M$. 
This consistency suggests that the velocity fields within MCs could inherit relative motion between clouds. 
Both short merger timescales and prevalence of velocity discontinuities 
support our proposed view of MCs as the transient processes of MCs characterized by slow mergers or splits 
among their `fundamental blocks' \citep{Yuan2022, Yuan2023, Yuan2023b, Yuan2024}.

In numerical simulations, the merger rates of clouds in the various galactic environments have been investigated.  
\cite{Tasker2009} estimate that the typical merger rate is around 25 Myr$^{-1}$, 
i.e. one merger per $\sim$ 1/5 of the orbital time, 
in a Milky-Way-like disk galaxy with a flat rotation curve but without spiral potential. 
In a hydrodynamic simulation of an isolated galaxy that 
incorporates heating and cooling of the interstellar medium (ISM), self-gravity, and stellar feedback,
cloud-cloud mergers occur every 8 - 10 Myr (about 1/15th of an orbit) with spiral arm, 
whereas they occur once every 28 Myr (1/5th of an orbit) without spiral arms \citep{Dobbs2015} .
When a spiral and bar potentials are imposed, 
merger rates can increase to one every 2 or 3 Myr (1 merger per 1/40th of an orbit) for massive clouds. 
In contrast, the majority of gravitationally bound but non-massive clouds experience 1 to 2 mergers during their lifetimes \citep{Fujimoto2014a}. 
Additionally, simulations by \cite{Skarbinski2023}, which include momentum and thermal energy from the supernovae, 
found that mergers occur at an average rate of 1-2 percent of clouds per Myr.
These mergers are primarily driven by the converging flows induced by galactic differential rotation (galatic shear) and 
stellar feedback. \cite{Jeffreson2021} revealed a scaling relation regarding the cloud merger rate, 
indicating that the merger rates of clouds are proportional to the crossing time between their centroids in a supersonically turbulent. 
Overall, these simulated results indicate that merger rates of MCs vary across different environments. 
Our observed $t_{\rm merge}$ is comparable to the findings from galaxies with the imposed spiral and bar potential. 
The scaling relation of \cite{Jeffreson2021} linking merger rates to centroid crossing times in turbulence, 
further validates our approach.

\subsection{The contribution of clouds merger to the MC constituents}

As discussed above, we roughly estimate the merger timescale of MCs and 
reveal that Type $L$ MCs experience multiple mergers within their crossing time. 
Additionly, the cloud-to-cloud velocity dispersion ($\sigma_{\rm cl}$) is higher for Type $S$ MCs ($\sim$ 7.6 km s$^{-1}$) 
than for Type $M$ MCs ($\sim$ 6.7 km s$^{-1}$), likely due to kinetic energy losses during mergers. 
These observations suggest that Type $L$ MCs may form through the hierarchical assembly of smaller MCs. 

To assess whether Type $L$ MCs primarily form from the coalescence of Type $S$ and $M$ MCs, 
we estimate the total gas mass in each type. 
Assuming a constant CO-to-H$_{2}$ conversion factor ($X_{\rm CO}$) and similar distances for MCs in the `Near' region, 
the integrated $^{12}$CO flux serves as a proxy for the molecular gas mass. 
The total integrated $^{12}$CO flux of Type $S$ MCs is 1.5 $\times$ 10$^{5}$ K km s$^{-1}$ arcmin$^{2}$. 
While for Type $M$ MCs, this value is 3.7 $\times$ 10$^{5}$ K km s$^{-1}$ arcmin$^{2}$, 
and for those in Type L, it is 1.5 $\times$ 10$^{6}$ K km s$^{-1}$ arcmin$^{2}$, as listed in Table \ref{tab:t_fac}. 
We find that the gas mass in Type $M$ MCs is roughly twice that in Type $S$, 
while Type $L$ MCs contain $\sim$ 10 times the mass of Type $S$. 
Over a timescale of $\sim$ 2 Myr (the crossing time for Type $M$ MCs),
the total mass produced by Type $S$ MCs is comparable to that in Type $M$.
Similarly, over $\sim$ 5 Myr (the crossing time for Type $L$ MCs), 
the combined mass from Type $S$ and $M$ MCs matches that in Type $L$,  
This indicates that the gas mass from MCs of Type $S$ and $M$ likely can account for most mass of Type $L$ MCs. 

In previous studies, MC complexes have been observed to be associated with Galactic supershells, 
such as HI supershells \citep{Dawson2011a, Dawson2011b} and three-dimensional C-shaped dust structures \citep{Edenhofer2024}.  
This suggests that the gathering process is likely associated with larger-scale dynamic processes, 
such as large-scale stellar feedback. Further analysis on the spatial distributions of larger-scale MCs and 
their relations with larger-scale dynamic activities in the extended regions is essential for validating that.

\section{Conclusions}

We mainly analyze 9617 $^{12}$CO molecular clouds (MCs) from the MWISP CO survey, 
which spans 450 deg$^{2}$ in Galactical longitude from 104$^{\circ}$.75 to 150$^{\circ}$.25, 
and latitude $|b| < 5^{\circ}.25$.
These MC samples primarily concentrate on the Local arm at a typical kinematical distance of $\sim$ 0.5 kpc. 
We measure their one-dimensional cloud-to-cloud velocity dispersions using two methods: 
(1) $\sigma_{\rm bin}$, the standard deviation of centroid velocities for MCs within spatial bins, 
and (2) $\sigma_{\rm bin, w}$, a flux-weighted analog that emphasizes larger clouds. 
We also categorize MCs by their sizes into three types: 
Type $S$ (0.15 -- 1.2 pc), Type $M$ (1.2 -- 4.8 pc), and Type L ($\gtrsim$ 4.8 pc), 
and present their spatial distributions.
Our key findings are: 

1. The typical cloud-to-cloud velocity dispersion across this Local arm segment 
is around 7.5$\pm$0.5 km s$^{-1}$ for $\sigma_{\rm bin}$ and 6.2$\pm$0.5 km s$^{-1}$ for $\sigma_{\rm bin, w}$. 

2. The cloud-to-cloud velocity dispersion between smaller MCs in Type $S$ 
($\sigma_{\rm bin}$ of 7.6$\pm$0.3 km s$^{-1}$ and $\sigma_{\rm bin, w}$ of 7.4$\pm$0.3 km s$^{-1}$) 
are systematically greater than those for larger clouds classified as Type $M$ 
($\sigma_{\rm bin}$ of 6.7$\pm$0.5 km s$^{-1}$ and $\sigma_{\rm bin, w}$ of 6.0$\pm$0.5 km s$^{-1}$). 

3. The spatial distribution of MCs in Type $S$ and $M$ is overall uniform when projected on the l-b plane. 

4. Mergers between MCs occur on timescales of 0.3 -- 0.9 Myr, 
shorter than their internal crossing timescales ($\sim$ 1 Myr for Type $S$, $\sim$ 2 Myr for Type $M$, 
and $\gtrsim$ 5 Myr for Type $L$.  
This suggests that mergers among MCs happen more rapidly than internal stabilization, 
especially for larger clouds (Type $L$).
This emphasis the dynamic, short-lived nature of molecular clouds, which are governed by frequent mergers.

\begin{acknowledgments}
    We are grateful for Yang Su, Min Fang, Fujun Du for their helpful discussion.
    This research was supported by the National Natural Science Foundation of China through grant 
    12041305 $\&$ 12303034 and the Natural Science Foundation of Jiangsu Province through grant BK20231104.
    This research made use of the data from the Milky Way Imaging Scroll Painting (MWISP) project, 
    which is a multi-line survey in $^{12}$CO/$^{13}$CO/C$^{18}$O along the northern galactic plane with PMO-13.7m telescope. 
    We are grateful to all of the members of the MWISP working group, particulaly the staff members at the PMO-13.7m telescope, 
    for their long-term support. MWISP was sponsored by the National Key R\&D Program of China with grant 2023YFA1608000 $\&$ 2017YFA0402701 and 
    the CAS Key Research Program of Frontier Sciences with grant QYZDJ-SSW-SLH047. 
\end{acknowledgments}

\software{Astropy \citep{astropy2013, astropy2018}, Scipy \citep{scipy2020}, Matplotlib \citep{Hunter2007}}
    
\clearpage
\appendix
\renewcommand{\thefigure}{\Alph{section}\arabic{figure}}
\renewcommand{\theHfigure}{\Alph{section}\arabic{figure}}
\setcounter{figure}{0}

\section{Observational Limitation}
Our MC samples are defined as continuous structures within the PPV space of the observed $^{12}$CO data, 
which are extracted using the DBSCAN algorithm. 
Consequently, these MC samples are likely influenced by the sensitivity and spatial resolution of the observed data, 
as well as the parameters used in the DBSCAN algorithm, i.e. $cutoff$, $\epsilon$, and $MinPts$. 
These factors may further impact our measurements of cloud-to-cloud velocity dispersion.  

The sensitivity of the observed data determines the completeness of the observed line fluxes, 
and the spatial resolution of observed data probably affect the observed brightness temperature, due to the beam dilution effect. 
\cite{Yan2021} used the MWISP CO data as a benchmark to simulate observations with larger beam sizes 
and higher noise levels, deriving corresponding values for the beam filling factor ($\eta$) and sensitivity clip factor ($\epsilon$). 
The sensitivity clip factor estimates the completeness of observed fluxes. 
The variations of $\eta$ and $\epsilon$ as beam sizes change are modeled by extrapolating an empirical function. 
For the larger-scale $^{12}$ CO maps, $\eta$ values are roughly equal to one for distances less than $\sim$ 6 kpc, 
but drop to $\sim$ 0.6 in the outer arm ($\sim$ 15 kpc). 
The $\epsilon$ shows a similar trend, however, they are systematically lower by $\sim$ 0.2.
For the extracted MC samples from the MWISP data, 
the $\eta$ values are close to one only for clouds with angular size greater than 10$^{\prime}$. 
In contrast, the values decrease to below $\sim$ 0.6 for MCs with angular sizes 
(after deconvolving the beam size) smaller than 4 arcmin.  
 
The sensitivity and beam filling factor significantly influence MC samples, 
particularly evident in the $cutoff$ values that establish their intensity thresholds.    
\cite{Yan2020, Yan2022} systematically examined how different DBSCAN parameters affect the extracted MC samples. 
As the $cutoff$ value increases, the total number, angular area, and integrated flux of MCs decrease, 
as illustrated in Figure 2 by \cite{Yan2022}. 
Specifically, smaller MCs may be undetected, the extracted areas of MCs can diminish, 
and some larger MCs may be split into smaller ones, as demonstrated in Figure 1 of \cite{Yan2022}. 
A $cutoff$=2$\sigma$ (where $\sigma$ is approximately 0.5 K) is employed to obtain 
the complete MC samples while preventing noise from being misidentified as fake clouds \citep{Yan2020, Yan2022}. 

\begin{figure*}[th]
    \centering
    \includegraphics[width=1.0\textwidth, interpolate=false]{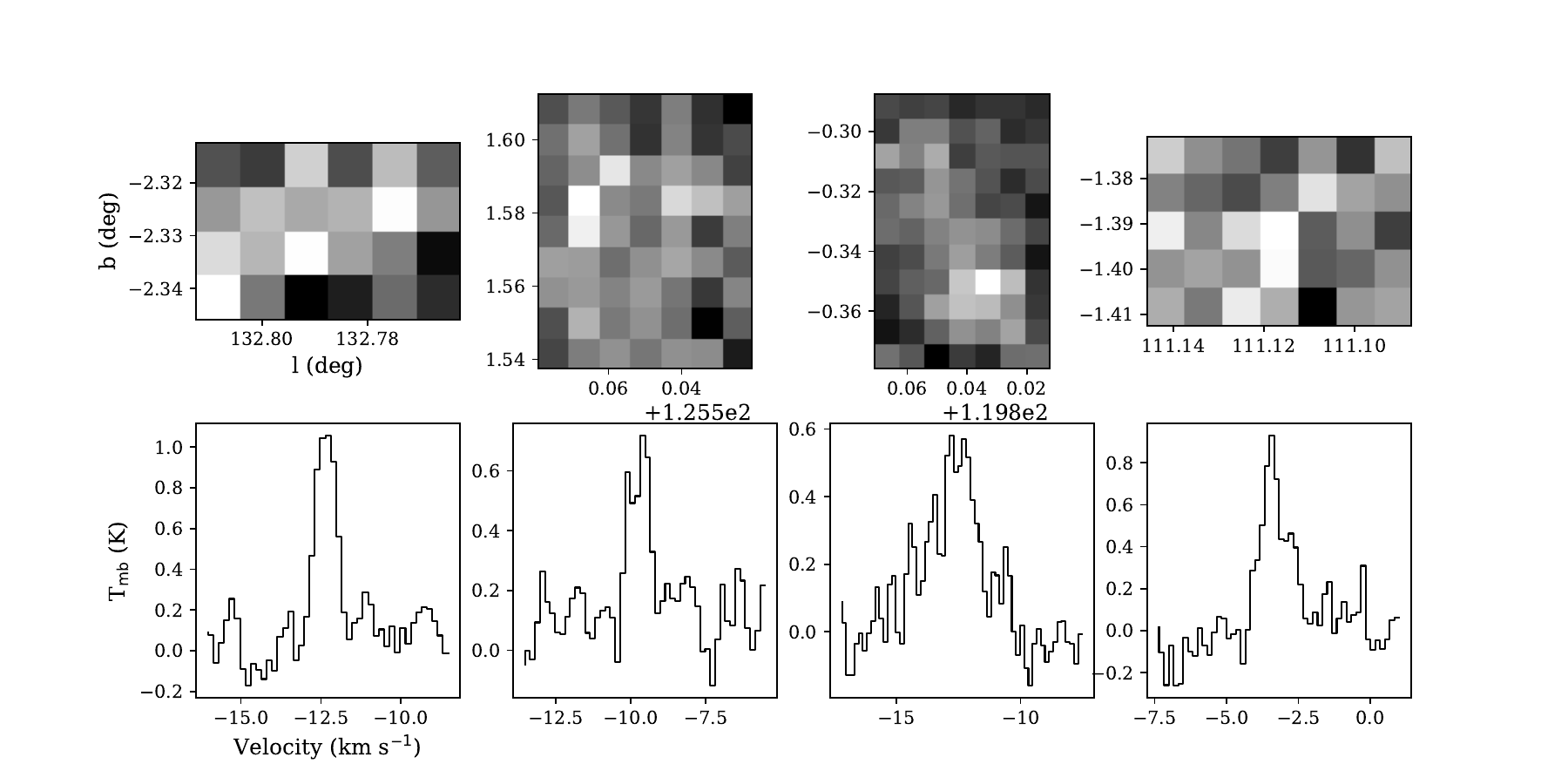}
    \includegraphics[width=1.0\textwidth, interpolate=false]{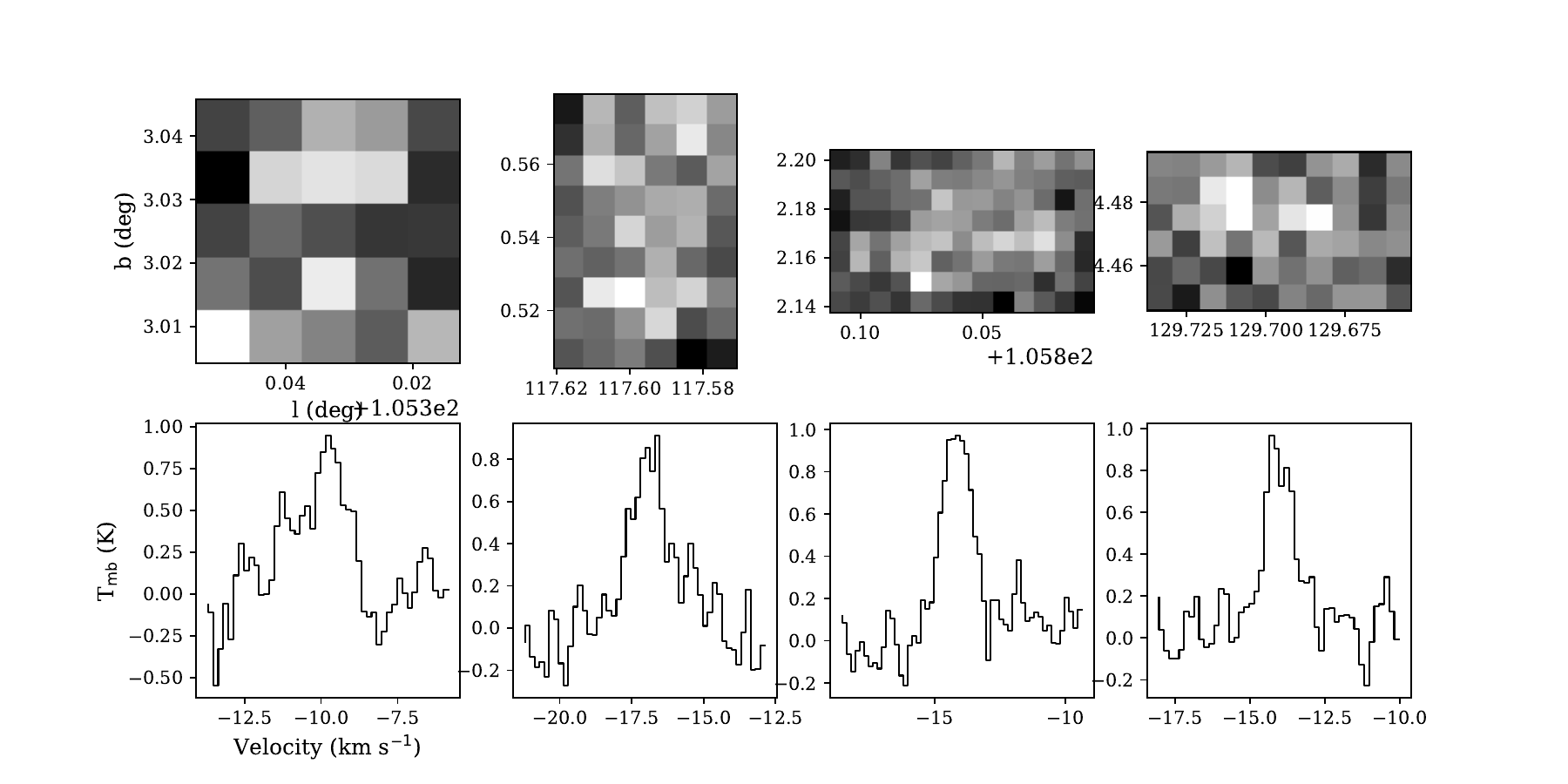}
    \caption{Example images of MCs, which are randomly selected from Type $S$ MCs with angular sizes ranging from 1 to 4 arcmin. 
    The upper images illustrate the velocity-integrated maps of $^{12}$CO lines for the MC samples, 
    the lower images display the corresponding averaged spectral lines of $^{12}$CO emission. \label{fig:cloudS_example}}
\end{figure*}

\section{The effect of different DBSCAN parameter values on the cloud-to-cloud velocity dispersion}

The DBSCAN algorithm utilizes three types of connectivity parameters, $\epsilon$ and $MinPts$, 
to define the neighborhoods in the PPV spaces. 
Connectivity 1 corresponds to $\epsilon$=1 in Euclidean distance, allowing for a maximum of 6 neighboring points. 
Connectivity 2 corresponds to $\epsilon$=$\sqrt{2}$, which permits up to a 18 neighboring points, 
while Connectivity 3 corresponds to $\epsilon$=$\sqrt{3}$, allowing for a maximum of 26 neighboring points. 
These connectivity types are illustrated in Figure 2 of \cite{Yan2020}.
In the studies of \cite{Yan2020, Yan2022}, 
they have systematically compared MC samples identified using three combinations of DBSCAN parameters.
Each combination corresponds to a specific connectivity type and an appropriate MinPts value: 
Connectivity 1 with MinPts = 4; Connectivity 2 with MinPts = 8; and Connectivity 3 with MinPts = 11.
The selection of MinPts is aimed at retaining as much $^{12}$CO emission fluxes as possible while filtering out noise. 
As shown in Figure 8 of \cite{Yan2022}, 
the differences among MC samples extracted using three DBSCAN parameter combinations 
are not significant. In particular, samples from Connectivity 2 and 3 are nearly identical. 
Meanwhile, Connectivity 1 tends to identify slightly more MCs, which is employed for the DBSCAN algorithm to extract our MC samples. 

\begin{figure*}[th]
    \centering
    \includegraphics[width=1.0\textwidth, trim=120 30 120 30, clip]{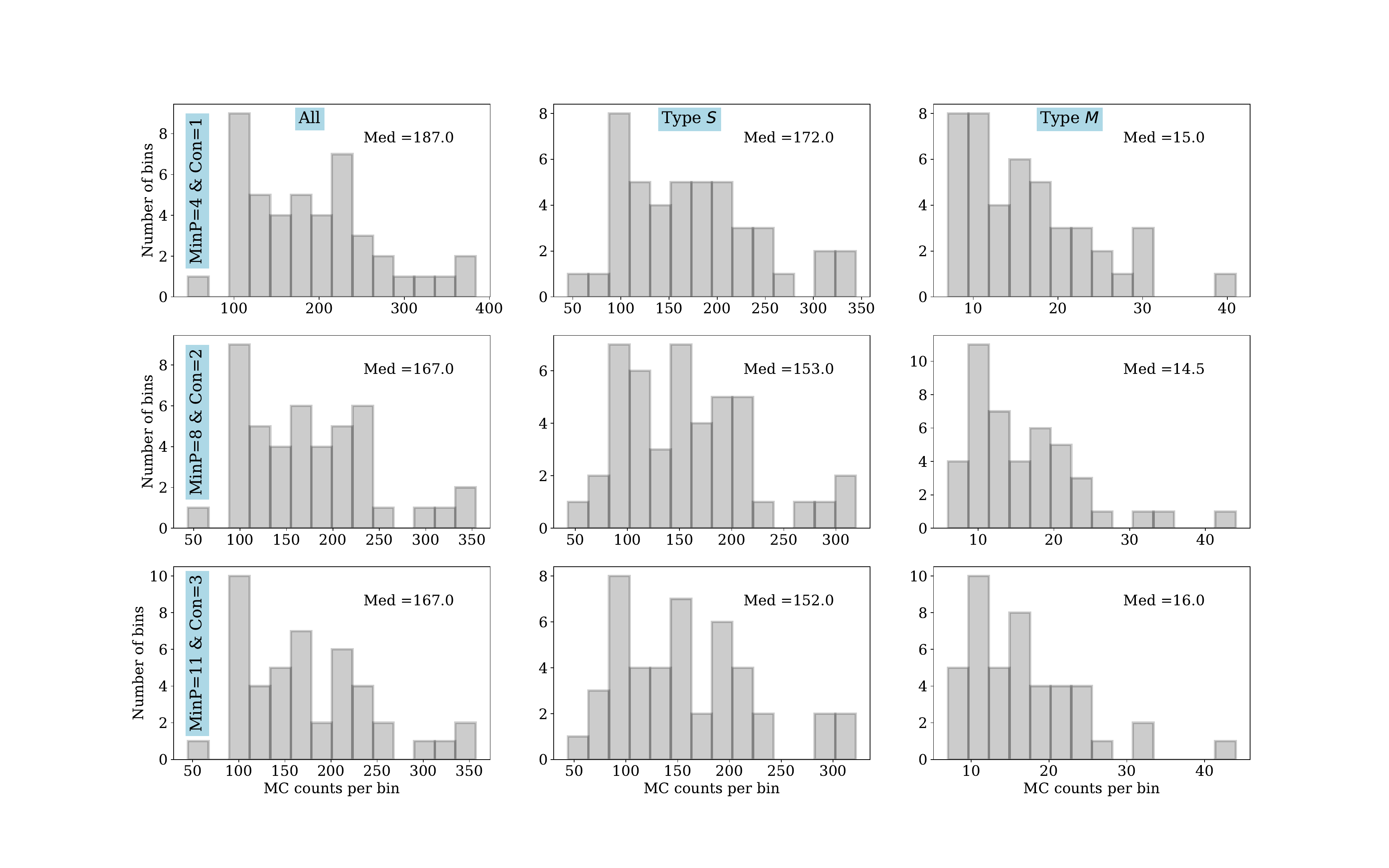} 
    \caption{The number distributions of MC samples in the `Near' region, which is counted within bins sized 3$^{\circ}$. 
    These MC samples are extracted using three connectivity parameters (i.e. Connectivity 1, 2, and 3) in the DBSCAN algorithm. 
    The `Con=1 $\&$ MinP=4' corresponds to Connectivity 1, the `Con=2 $\&$ MinP=8' corresponds to Connectivity 2, 
    and the `Con=3 $\&$ MinP=11' corresponds to Connectivity 3. In each category of MC samples, 
    MCs with angular sizes between 1 and 8 arcmin are classified as type $S$, those with angular sizes between 8 and 32 arcmin 
    are classified as type $M$, and those greater than 32 arcmin are classified as type $L$. 
    The median values for each number distribution are indicated in the corresponding panel. \label{fig:cloudN_dbscan}}
\end{figure*}

To investigate the impact of connectivity parameters of the DBSCAN algorithm on 
the measurement of cloud-to-cloud velocity dispersions, 
we also utilized the DBSCAN parameters of Connectivity 2 and 3 to extract MC samples. 
We then calculated the corresponding $\sigma_{\rm bin}$ and $\sigma_{\rm bin, w}$ values within a uniform bin size of 3$^{\circ}$. 
Figure \ref{fig:cloudN_dbscan} displays the distribution of cloud numbers within each bin across the three MC samples. 
We found that the number of MCs in each bin extracted using Connectivity 1 is greater than that from Connectivity 2 and 3. 
The statistical differences are about 20 and primarily stem from the MCs categorized as type $S$. 
In contrast, for MCs classified as type $M$ across the three samples, the number of MCs in each bin is fairly consistent. 

\begin{figure*}[th]
    \centering
    \includegraphics[width=1.0\textwidth, trim=120 30 120 30, clip]{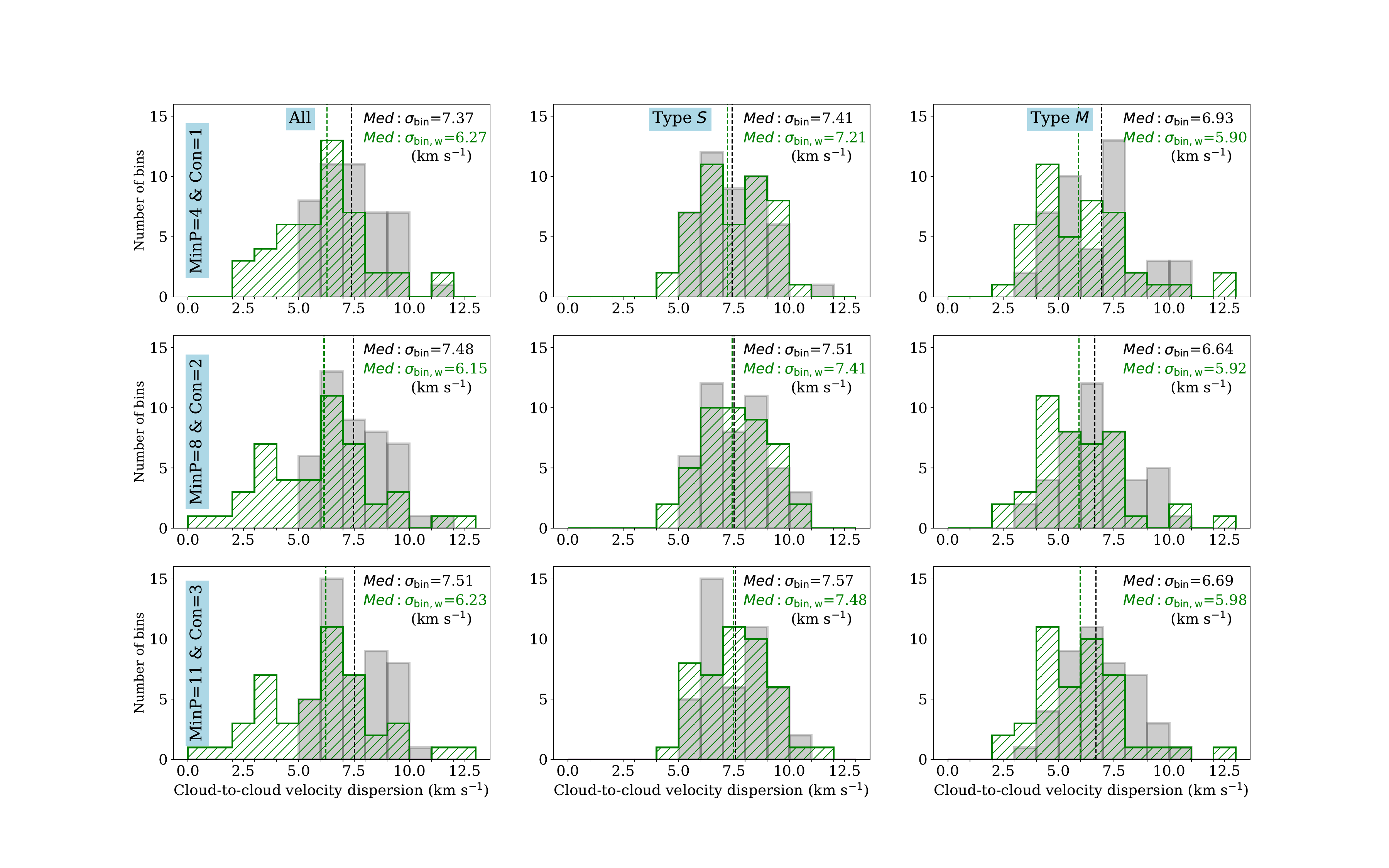} 
    \caption{Same MCs samples as illustrated in Figure \ref{fig:cloudN_dbscan}, 
    but for histogram distributions of their cloud-to-cloud velocity dispersions of $\sigma_{\rm bin}$ (gray) and $\sigma_{\rm bin, w}$ (blue) for, 
    which is calculated at bins sized 3$^{\circ}$. The median values of $\sigma_{\rm bin}$ and $\sigma_{\rm bin, w}$ are indicated 
    in the corresponding panel. \label{fig:vsigma_dbscan}}
\end{figure*}

Figure \ref{fig:vsigma_dbscan} illustrates the distributions of $\sigma_{\rm bin}$ and $\sigma_{\rm bin, w}$ from three 
MC samples. For the entire MC samples, the median values of $\sigma_{\rm bin}$ and $\sigma_{\rm bin, w}$ 
are close to $\sim$ 7.5 km s$^{-1}$ and $\sim$ 6.2 km s$^{-1}$, respectively. 
Their statistical differences between MC samples extracted from three connectivity types 
are less than 0.1 km s$^{-1}$.
For MCs in type $S$, the median values of $\sigma_{\rm bin}$ and $\sigma_{\rm bin, w}$ 
are also around 7.5 km s$^{-1}$ and 7.4 km s$^{-1}$, respectively, 
with differences in statistics among three MC samples being less than 0.2 km s$^{-1}$.
In the case of MCs in type $M$, the median value of $\sigma_{\rm bin}$ is about 6.9 km s$^{-1}$ for 
those extracted using Connectivity 1, 
while for those from Connectivity 2 and 3, it is around 6.7 km s$^{-1}$. 
However, for $\sigma_{\rm bin, w}$, the median value is around 5.9 km s$^{-1}$, 
with differences between the three MC samples being less than 0.1 km s$^{-1}$.
Overall, our measurements of $\sigma_{\rm bin}$ and $\sigma_{\rm bin, w}$ values 
are not significantly affected by the connectivity types used in the DBSCAN algorithm, 
with the differences remaining below 0.2 km s$^{-1}$.

\clearpage


\end{document}